\documentclass[aps,prx,showpacs,twocolumn,reprint,superscriptaddress]{revtex4-2}

\usepackage{qcircuit}
\usepackage[dvips]{graphicx}
\usepackage{amsmath,amssymb,amsthm,mathrsfs,amsfonts,dsfont,mathtools,physics,stmaryrd}
\usepackage{epsfig}
\usepackage{braket}
\usepackage{hyperref}
\usepackage{bm}
\usepackage{enumerate}
\usepackage[T1]{fontenc}
\usepackage[dvipsnames]{xcolor}
\usepackage{graphicx}
\hypersetup{
    colorlinks=true,
    linkcolor=Tan,
    filecolor=BrickRed,      
    urlcolor=BrickRed,
    citecolor=BrickRed
}

\usepackage{enumitem}
\usepackage[capitalize]{cleveref}
\usepackage[normalem]{ulem}
\usepackage{comment}
\usepackage{xcolor}

\usepackage{amsthm}

\Crefname{theorem}{Theorem}{Theorems}
\theoremstyle{remark}

\renewcommand{\tr}{\mathrm{Tr}}

\newcommand{\qmaddress}{\affiliation{Quantum Motion, 9 Sterling Way, London N7 9HJ, United Kingdom}}
\newcommand{\oxddress}{\affiliation{Department of Materials, University of Oxford, Parks Road, Oxford OX1 3PH, United Kingdom}}
\newcommand{\cicddress}{\affiliation{CIC nanoGUNE Consolider, Tolosa Hiribidea 76, E-20018 Donostia-San Sebastian, Spain}}
\newcommand{\ikddress}{\affiliation{IKERBASQUE, Basque Foundation for Science, E-48011 Bilbao, Spain}}

\begin{document}

\title{Harnessing electron motion for global spin qubit control}
\author{Hamza Jnane}
\thanks{These authors contributed equally \\ Corresponding email: hamza@quantummotion.tech}
\qmaddress

\oxddress
\author{Adam Siegel}
\thanks{These authors contributed equally \\ Corresponding email: hamza@quantummotion.tech}
\qmaddress
\oxddress

\author{M. Fernando Gonzalez-Zalba}
\qmaddress
\cicddress
\ikddress

\begin{abstract}
Silicon spin qubits are promising candidates for building scalable quantum computers due to their nanometre scale features. However, delivering microwave control signals locally to each qubit poses a challenge and instead methods that utilise global control fields have been proposed. These require tuning the frequency of selected qubits into resonance with a global field while detuning the rest to avoid crosstalk. Common frequency tuning methods, such as electric-field-induced Stark shift, are insufficient to cover the frequency variability across large arrays of qubits. Here, we argue that electron motion, and especially the recently demonstrated high-fidelity shuttling, can be leveraged to enhance frequency tunability. Our conclusions are supported by numerical simulations proving its efficiency on concrete architectures such as a 2$\times$N array of qubits and the recently introduced looped pipeline architecture. Specifically, we show that the use of our schemes enables single-qubit fidelity improvements up to a factor of 100 compared to the state-of-the-art. Finally, we show that our scheme can naturally be extended to perform two-qubit gates globally.
\end{abstract}

\maketitle

\section{Introduction}

Silicon-based spin qubits in semiconductor quantum dots are a promising platform due to their small footprint, high control fidelity~\cite{mills_2022, noiri_2022,xue_2022}, long coherence times \cite{stano_review_perfs_2022}, and their compatibility with nanofabrication techniques of the semiconductor industry~\cite{Maurand2016AQubit,gonzalez_zalba_scaling_2021,zwerver_qubits_made_2022,chittockwood2024exchangecontrolmosdouble, steinacker2024300mmfoundrysilicon, George2025}. For spin qubits to be a viable option for building scalable quantum computers, methods for controlling a large number of qubits -- potentially millions to implement error correction -- need to be developed. Particularly, when encoding a qubit in the spin of a single electron, a microwave field in resonance with the Zeeman-split levels leads to Rabi oscillations of the spin system. These fields can be applied locally through surface gate electrodes, but this approach, applied at scale, presents formidable challenges for signal delivery. Instead, a global control field can be sent and selected spins can be brought in and out of resonance by means of frequency tuning~\cite{kane1998,vahapoglu_single-electron_2021, Hansen_2024, Fayyaz_2023}. Unfortunately, common frequency-tuning techniques are unable to fully correct for the spread of qubit frequencies that are typically found in devices due to variations in the local spin-orbit fields, which for electron spins in silicon can be of the order of 1\%~\cite{Veldhorst2015,Huang2017, li_crossbar_2018,ferdous_interface_2018}. 

To ameliorate the situation, recent studies have shown that the spin qubit frequency spread can be reduced by choosing the angle of the static magnetic field with respect to the crystallographic axes of the sample~\cite{Tanttu2019, Hansen_2024}. Complementarily, one can tune the qubit frequencies using electric-field-induced Stark shifts, but its limited tuning range -- one order of magnitude smaller than the typical frequency spread $\Delta g/g_0 \lesssim 1\%$ -- forces further measures~\cite{Ferdous2018}. A partial solution, known as the \textit{binning method}, proposed tuning the $g$-factors using Stark shifts to place the qubits into a number of distinct-frequency bins thereby mitigating the effect of frequency crowding. In this way, the minimum frequency spacing does not decrease with system size \cite{patomäki2023pipeline, Fayyaz_2023}. By sending tones at resonance with each bin, selected qubits can be driven and the desired global single-qubit gates can be performed. Reference \cite{Fayyaz_2023} extended this scheme and suggested a way to drive not all but a single qubit, by adding SWAP gates. However, none of these works allows for the reliable implementation of single-qubit gates on any subset of qubits, without halving the already critical bin spacing.

Here, we propose solving the challenge of the limited frequency tuning range by leveraging electron motion through the spin qubit device. Namely, we suggest that the qubit frequencies can be better homogenised by spin shuttling (or alternatively via exchange-driven two-qubit SWAPs) . More specifically, if an electron experiences a collection of $g$-factors at a rate that is sufficiently fast compared to the Rabi frequency and the dispersion of the qubit frequencies, the electron will acquire an effective homogenised $g$-factor equal to the mean of all experienced $g$-factors. With this approach, the frequency tunability can now be of the order of the natural frequency spread rather than just a fraction as is typically the case for Stark shift tuning. While both methods, shuttling and exchange, leverage the same simple idea, we find that the shuttling-based homogenisation is more scalable, and can additionally be utilised for the implementation of two-qubit gates. The frequency homogenisation technique enables driving arrays of qubits accurately by using just a single global drive tone as opposed to tens of frequencies for the binning method, substantially simplifying the microwave control electronics. 

More than a key component for the development of fault-tolerant architectures \cite{xue_qubus_2024, adam_two_by_N_2024, siegel_snakes_2025, li_crossbar_2018, Cai_2023}, spin shuttling turns out to be an enabling tool in the search for homogeneous devices. It is worth mentioning that our idea is reminiscent of the frequency narrowing phenomenon \cite{Langrock_2023}; however, this concept was never used to perform quantum gates. In this work, we concentrate on electron spin qubits in silicon because of their small spin-orbit fields, meaning the spin quantisation axis is independent of electron's location in the device. However, as we shall see later, our work can be extended to systems subject to strong spin-orbit coupling that present large variations in the spin quantisation axis across the device.

The paper is organised as follows. In \cref{sec: preliminary}, we introduce our model for the qubits. In \cref{sec:general_idea}, we abstractly describe the homogenisation principle before giving two methods to physically implement it, using exchange coupling or spin shuttling. Focusing on the more promising shuttling-based protocol, we then explore its performance for two architectures, namely the 2$\times$N \cite{adam_two_by_N_2024, micciche_optimizing_2xN_2025} and looped pipeline architectures \cite{Cai_2023} in \cref{sec:applications}. Finally, we present our conclusions in \cref{sec:conclusion}.

\section{\label{sec: preliminary} System Hamiltonian}

Our system can be described by the following Hamiltonian:
\begin{equation} \label{eq:hamiltonian}
    H = \sum_{<i,j>}J_{i,j} \vec{S}_i . \vec{S}_{j} + \sum_{i=1}^{n_q} g_i \vec{S}_{i} . \vec{B}
\end{equation}
where $n_q$ is the total number of qubits, $J_{i,j}$ is the tunable exchange interaction between neighbouring spins $\vec{S}_i$ and $\vec{S}_{j}$ and $g_i$ are the electron $g$-factors. We set $\hbar=\mu_B=1$, thus identifying energies, frequencies and magnetic fields. The exchange coupling strength can be tuned in-situ via electric fields and whose typical on-state value ranges between 10 and 1000 MHz \cite{xue_2022, mills_2022, noiri_2022, sheahta_modelling_2023, cifuentes_path_integral_exchange_2023, jnane_ab_initio_exchange_2024}. The field $\vec{B} = \sum_j \vec{B}_j$ is a global magnetic field, in the sense that it is applied across the entire device. It is composed of a single static component $B_z=B_0$, and an oscillatory component $B_x=B_1\sum_j\cos(\omega_j t)$. Typical experimental values for $B_0$ range from 0.1 T to 1.4 T. As we shall see later, for our frequency homogenisation protocol, it is favourable to use values from the low end of the range to reduce the qubit energy spread. However, this complicates the task of single-qubit addressability owing to reduced frequency spread, which justifies the design of more advanced schemes for handling frequency crowding. We therefore set $B_0=0.1$ T in the rest of this paper unless otherwise stated. The oscillatory component of the drive will lead to magnetic dipole transitions with an amplitude $\Omega$, whose value is usually around 1 to 5 MHz \cite{yoneda_2018, yang_single_q_2019, xue_2022,mills_2022, noiri_2022}. Additionally, the qubits can be shuttled along the dots at a speed $v$ between 10 and 50 m/s~\cite{Yoneda_2021, Seidler_2022, de_smet_shuttling_2024}. We further assume an interdot spacing $d=100$~nm, resulting in a shuttling frequency $v/d=100-500$~MHz. As for the $g$-factors, $g_i$, modelling their distribution is complex and it is currently an active field of research. We elaborate on this matter in \cref{app:model_g_factor}. Note that in the common case of a single excitation frequency, the Hamiltonian of \cref{eq:hamiltonian} is made time-independent by studying it in the rotating frame at the frequency of the drive. However, it is not possible to perform such transformation here as we will be driving the spins with multiple tones $\omega_j$.

We note that our system is most accurately described by considering the full $g$-tensor in \cref{eq:hamiltonian}, which can effectively change the axes along which the effective static and oscillatory magnetic fields are applied \cite{cifuentes_bounds_2024}. We show in \cref{app:g_tensor} that this has minimal impact on the fidelity of a single-qubit gate for electron spins in silicon, even when the static field includes components along the $x$ and $y$ axes. Therefore, in the rest of the paper, we will assume that the static and oscillatory fields are indeed applied along the $z$ and $x$ axes, respectively, and that $g_i$ are scalars. However, our scheme could be applied to more general species of spin qubits and semiconducting hosts. In these cases -- where spin-orbit coupling effects are stronger -- variations of the quantisation axes with the electron motion induce unwanted unitary gates. Fortunately, these could be calibrated away provided prior knowledge of the $g$-tensor~\cite{Wang2024}. We leave this study for future work and focus here on electron spins in silicon.

In \cref{tab:parameters}, we summarise the parameters used in this study in decreasing order of magnitude considering $B_0 = 0.1$ T. The apparatus time resolution $\tau_r$ is added, as an upper (frequency) limit for all other parameters. The ratio between these parameters will be relevant for the characterisation of the performance of our scheme.

\begin{table}[]
    \centering
    \begin{tabular}{ |l|c|c|}
     \hline
      Parameter & Notation & Order of magnitude \\
     \hline
      Time resolution & $1/\tau_r$ & 10 GHz \\
      Tunnel coupling & $t_c$ & 1-10 GHz \\
      Zeeman splitting (Avg) & $\omega_q/2\pi$ & 2.8 GHz \\
      Shuttling speed & $v/d$ & 100-500 MHz \\
      Exchange coupling & $J/2\pi$ & 10-1000 MHz \\
      Zeeman splitting (Std) & $\Delta\omega_q/2\pi$ & 2.8-28 MHz \\
      Drive strength & $\Omega/2\pi$ & 1-5 MHz \\
      Stark-shift tunability & $\delta\omega_q/2\pi$ & 0.28-2.8 MHz \\
     \hline
    \end{tabular}
    \caption{Relevant parameters and frequency scaling at $B_0=0.1$~T.}
    \label{tab:parameters}
\end{table}

\section{\label{sec:general_idea} Motion-based single-spin control}

If an electron explores a collection of $g$-factors at a rate that is much faster than the frequency of a quantum gate and the qubit's frequency dispersion, its $g$-factor will effectively be homogenised. Such a process facilitates the implementation of global gates by sending a single driving tone to a set of distinct target qubits.

\begin{figure}
    \centering
    \includegraphics[width=\linewidth]{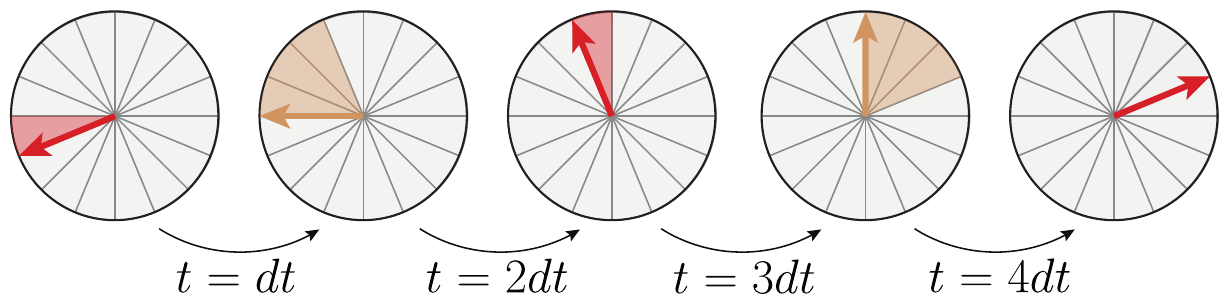}
    \caption{Schematic time-evolution of a single qubit oscillating between two quantum dots with distinct frequencies $\omega_{q,1}$ (red) and $\omega_{q,2}$ (yellow). We assume that the transfer rate between the dots is fast enough that the electron instantaneously swaps position after each time step. Assuming that it starts in the first dot, the qubit rotates by one tile and ends up in the second dot after one time step. Next, it rotates by three tiles and comes back to the first dot. Finally at $t=4dt$, the qubit is back to its original position and has covered eight tiles. This leads to the conclusion that the resulting dynamics is a precession of the qubit at a frequency $\bar{\omega}_q = (\omega_{q,1}+\omega_{q,2})/2$.}
    \label{fig:schematic_time_evolution}
\end{figure}

To provide a simple example, consider a single electron trapped in a double quantum dot (DQD). Suppose that the left and right dots have $g$-factors $g_1$ and $g_2$, respectively. In the laboratory frame of reference, the qubit would precess with a different Larmor frequency (set by the $g$-factor) depending on its location. Now suppose that the qubit can instantaneously be transferred from one dot to the other at every time step. The resulting time dynamics is schematically represented in \cref{fig:schematic_time_evolution}. One can see that, because of the transfer, the qubit now rotates with an average frequency $\bar{\omega}_q = (\omega_{q,1}+\omega_{q,2})/2$. This leads to the conclusion that the qubit now has a modified resonant frequency equal to $\bar{\omega}_q$. Numerical simulations in the next section in the presence of a drive will support this hypothesis for finite transfer rates. Following the above intuition, we report two different physical mechanisms (see \cref{fig:physical_implementation}) that could be leveraged to homogenise $g$-factors and hence qubit frequencies.

\subsection{Exchange-based homogenisation}
\subsubsection{Single-qubit gates on two electrons spins} 
\begin{figure}
    \centering
    \includegraphics[width=\linewidth]{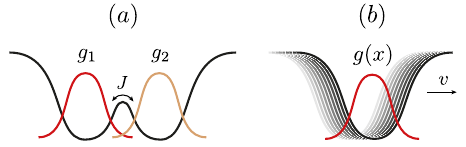}
    \caption{Physical implementations of the motion-based $g$-factor homogenisation. In (a), we consider two quantum dots with respective $g$-factors $g_1$ and $g_2$. The exchange coupling $J$ is used to homogenise the frequencies of the two electrons. This allows to reduce the number of distinct frequencies in the spectrum and thus reduce crosstalk. In (b) we use a completely different method in which we physically move the electron to perform the homogenisation. Here, the $g$-factor evolves continuously along the shuttling path. We find that this method is the most powerful as it can be generalised to an arbitrary number of qubits and scales better.}
    \label{fig:physical_implementation}
\end{figure}

One way to implement the example presented in \cref{fig:schematic_time_evolution} is to make use of the exchange interaction as shown in \cref{fig:physical_implementation}(a), where two separated spins in a DQD are subject to exchange interactions. The information transfer rate here corresponds to the exchange coupling $J$. By turning on $J$ well above their Zeeman energy difference, spins in neighbouring quantum dots start to swap information.

\begin{figure}
    \centering
    \includegraphics[width=\linewidth]{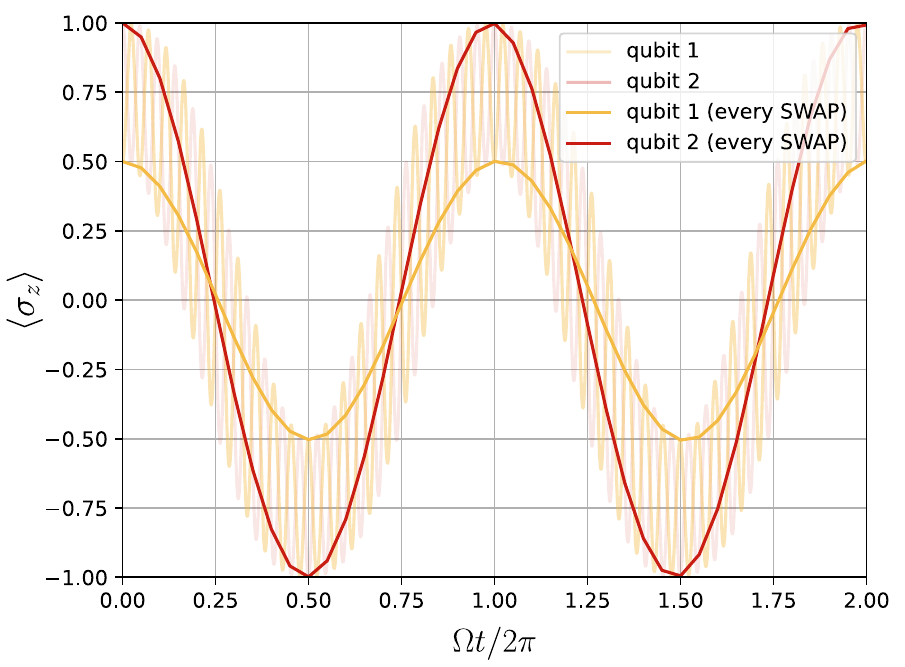}
    \caption{Time evolution of $\langle\sigma_z\otimes I\rangle$ and $\langle I\otimes\sigma_z\rangle$ under the two-qubit Hamiltonian of \cref{eq:hamiltonian}. We set $\omega_q^{12}=\Omega$ and $J=20 \, \omega_q^{12}$ where $\omega_q^{12} = (\omega_{q,2}-\omega_{q,1})$. The initial state is given in \cref{eq:init_state_swap}. The faded lines correspond to the full time evolution, while the solid lines are obtained from sampling the full data every $2\pi/J$.}
    \label{fig:time_evolution_2_qubits}
\end{figure}

Let us add an oscillatory magnetic field with frequency $\omega= \bar{\omega}_q$. If the intuition exposed previously is right, one would expect to perform a perfect $X$-rotation on both qubits. This is confirmed by the numerics of \cref{fig:time_evolution_2_qubits}. Starting from the arbitrary initial state,
\begin{align} 
    \ket{\psi_0} = \ket{0} \otimes \left(\sqrt{\frac{3}{4}}\ket{0} + \sqrt{\frac{1}{4}}\ket{1}\right),
    \label{eq:init_state_swap}
\end{align}
we plot the expectation value of $\sigma_z$ over time for both qubits, at a finite $J=20 \, \omega_q^{12}$ with $\omega_q^{12} = (\omega_{q,2}-\omega_{q,1})$, additionally choosing $\omega_q^{12}=\Omega$. These simulations are simply obtained by solving Schrödinger equation by time discretisation, using 10,000 time steps. The red and yellow faded lines respectively represent the full time-evolution of $\bra{\psi(t)}\sigma_z\otimes I\ket{\psi(t)}$ and
$\bra{\psi(t)}I\otimes\sigma_z\ket{\psi(t)}$. These feature fast oscillations at a frequency $J$, which corresponds to the swapping of the qubits, and slower oscillations at frequency $\omega=\bar{\omega}_q$, which are triggered by the driving field. When sampling points from the red and yellow faded data every $2\pi/J$, one obtains the solid lines, which qualitatively look like an $X$ gate on both qubits. This means that, at finite but large enough $J$, the individual qubits indeed acquire the same effective frequency $\bar{\omega}_q$ and are thus resonantly driven by a single drive at $\omega=\bar{\omega}_q$. 
Unfortunately, the exchange interaction can entangle the qubits. Therefore, in order to implement an $X$ gate on both qubits without any additional entanglement creation, one must properly tune $J$ such that the qubits go back to their original positions in the time taken by the $X$ gates. This condition reads:
\begin{equation*}
    \frac{2p\pi}{J} = \frac{\pi}{\Omega}, ~~ p\in\mathbb{N}
\end{equation*}
or equivalently:
\begin{equation} \label{eq:integer_J}
    J = 2p\Omega, ~~ p\in\mathbb{N}.
\end{equation}

In order to further estimate the similarity of the implemented gate $U$ to the target unitary $U_{\text{target}}=X\otimes X$, one can evaluate the average single-qubit fidelity $\mathcal{F}$, defined as 
\begin{align}
    \mathcal{F} = \frac{1}{4}\left|\tr(U_{\text{target}}^\dagger \tilde{U})\right|^{2/n_q}
    \label{eq:fidelity}
\end{align}
with $n_q = 2$ and $\tilde{U}$ is a slightly modified version of the time evolution operator $U$ (where we removed \emph{known but unwanted} $Z$ rotations) as explained in \cref{app:numerical_simulations}. This quantity is dimensionless and thus only depends on the ratios of the three parameters dictating the dynamics of the two-qubit system: $J$, $\Omega$ and $\omega_q^{12}$. Only the difference between $\omega_{q,1}$ and $\omega_{q,2}$ matters, as their absolute values can be absorbed in a change of frame. 

\begin{figure}
    \centering
    \includegraphics[width=\linewidth]{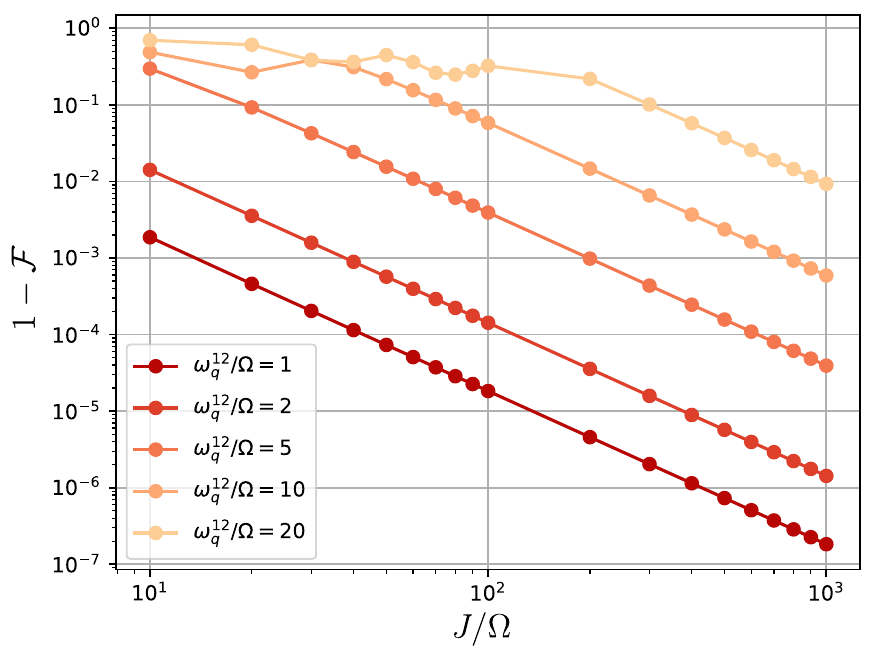}
    \caption{Infidelity of the swapping protocol. It is computed as the infidelity of implementing an $X\otimes X$ gate when driving two continuously-swapped qubits with frequencies $\omega_{q,1}$ and $\omega_{q,2}$ with a single driving tone at $\omega=\bar{\omega}_q$. The data is plotted against $J/\Omega$ and $\omega_q^{12}/\Omega$, where $J$ is the exchange coupling and $\omega_q^{12}=\omega_{q,2}-\omega_{q,1}$. $J$ is always chosen such that the condition in \cref{eq:integer_J} is satisfied.}
    \label{fig:swap_2_qubits}
\end{figure}

If the condition in \cref{eq:integer_J} is always enforced, one can expect the infidelity $1-\mathcal{F}$ to decrease to 0 when $J$ increases: this is confirmed by the numerics of \cref{fig:swap_2_qubits}. One can also observe that when $\omega_q^{12}$ is decreased, the infidelity also decreases: this is expected as the limiting case of $\omega_q^{12}=0$ should produce null infidelity. Now, using example values from \cref{tab:parameters}, such as $J/\Omega=100$ and $5 \gtrsim\omega_q^{12}/\Omega \gtrsim 1$, simulations produce an infidelity between $0.001\%$ and $1\%$. To understand what limits the fidelity, we explore the Schmidt decomposition of $U$ in \cref{app:schmidt}, and find that at finite transfer speed $J$, the time evolution operator $U$ is indeed creating entanglement between the qubits, and cannot be decomposed into a product of two single-qubit gates.

As noted previously, the frequency homogenisation efficiency improves as one uses larger transfer rates. With the exchange coupling, one is typically limited to transfer rates of 0.01-1~GHz, in practice. However, by removing one electron from the DQD, one could instead leverage the tunnel coupling $t_c$ for electron transfer. This quantity can typically reach 1-10 GHz \cite{Croot_2020}, which should lead to faster transfer rates and hence even better global control fidelities. Moreover, contrary to the exchange-based homogenisation, this method does not create any electron-electron entanglement, which is a limiting factor to our fidelity. We investigate this alternative option in \cref{app:tunnnel_coupling}.

\subsubsection{Single-qubit gates on more than two electrons}
The extension of the SWAP-based scheme to more than two qubits is not trivial. Indeed, one would naively think that turning on the $J$-coupling between all adjacent dots would produce the same desired outcome. Unfortunately, multi-qubit SWAP gates are not implemented by simply turning on all $J$-couplings. Rather, one could think of implementing a multi-qubit SWAP gate by decomposing it into two-qubit SWAPs, which could be performed by turning the $J$-couplings on and off in a sequential fashion. If this method theoretically works, it is however experimentally challenging. 

One of the most significant sources of errors in implementing SWAP gates is charge noise, which arises from two-level fluctuators (TLFs) \cite{kepa_2023, yoneda_noise_correlation_2023, rojas_spatial_correlations_2023}. These TLFs lead to unknown modifications in the exchange coupling \cite{jnane_ab_initio_exchange_2024, cifuentes_path_integral_exchange_2023, sheahta_modelling_2023}, leading to timing errors that reduce the fidelity. 
When decomposing a multi-qubit SWAP gate into two-qubit SWAPs, a large number of sequential SWAPs between different qubits are required. This results in an exponential decrease in overall fidelity with the number of SWAPs, rendering the protocol impractical. However, in the two-qubit case, we only need to turn the exchange coupling on and off once, so the timing error only affects the process a single time. Moreover, since the exchange is active for an order of microseconds, charge fluctuations on this timescale should average out the variations of the exchange coupling, leading to higher fidelity \cite{jnane_ab_initio_exchange_2024}.
Finally, it is important to note that as charge noise modifies the exchange coupling $J$, it will lead to a violation of the condition in \cref{eq:integer_J}. As this paper focuses on addressability, the impact of charge noise on our scheme is left for further study.

\subsection{\label{subsec:shuttling} Shuttling-based homogenisation}
\subsubsection{Single-qubit gate on a single electron}
In the previous section, we focused on a mechanism that allows for a qubit to oscillate between two regions of fixed $g$-factors. In this section, we advocate for the use of a more powerful mechanism, which leads to a broader exploration of the $g$-factor landscape: electron shuttling. Here, the electron is physically moved thanks to the electrical control of the trapping potential and can thus explore a continuous landscape of $g$-factors as illustrated in \cref{fig:physical_implementation}(b). Moreover, this scheme can easily be generalised to more than two qubits, as multiple electrons can be shuttled at the same time without any entanglement being generated between them.

There are two main ways to perform electron shuttling: the bucket brigade mode \cite{Yoneda_2021, buonacorsi_simulated_shuttling_2020, ginzel_spin_shuttling_dqd_202, mills_shuttling_2019}, in which the electron is moved from one dot to another by altering the detuning between them and the conveyor-belt mode \cite{Seidler_2022, xue_qubus_2024, Langrock_2023, kunne_spinbus_2024, de_smet_shuttling_2024, minjun_shuttling_2024, nagai_digital_controlled_2025}, in which it is the trapping potential itself that moves, carrying the electron with it. Here, we will focus on the latter as it is more advantageous in terms of scalability \cite{xue_qubus_2024, de_smet_shuttling_2024}. Recent experiments showed that coherent spin shuttling in Si/SiGe can be achieved with fidelities as high as $99.99\%$ for gate-to-gate distance and around $99 \%$ for a distance of 10 $\mu$m \cite{de_smet_shuttling_2024}. As in the previous sections, an electron shuttled in a region of varying $g$-factor will acquire an average effective frequency if shuttled fast enough, with the difference that we are now considering a continuously varying $g$ (see \cref{eq:avg_g_factor}). This frequency narrowing phenomenon had in fact already been observed in the context of shuttling, but had never been used for single-qubit addressability \cite{Langrock_2023}. 

More precisely, during the gate time $T = \pi/\Omega$, an electron with trajectory $x(t)$ will obtain the homogenised $g$-factor $\bar{g}$ given by, 
\begin{align}
    \bar{g} = \frac{1}{T}\int_{0}^{T}g(x(t))dt.
    \label{eq:g_bar_shuttling}
\end{align}
Contrary to the exchange- and tunnel-based protocols, it seems that we now require a perfect knowledge of $g(x(t))$ at each time $t$ in order to find the drive frequency. While this would be quite challenging, one can rather directly determine the drive frequency by shuttling the electron along $x(t)$ and by sweeping the frequency of the drive $\omega$ until resonance with the qubit is obtained \cite{koppens_driven_2006}. One can then deduce $\bar{g}$ from $\omega$ as $\omega = \bar{g}B_0$. 

However, depending on the shuttling trajectory $x(t)$, such characterisation might not be needed. Suppose that we let the electron explore dots without repetitions, \textit{e.g.} by shuttling it on a straight line for a time $T$. As the travelled distance increases, the homogenised $g$-factor $\bar{g}$ given by \cref{eq:g_bar_shuttling} will tend towards the device average $g$-factor, $g_0$.
For a sufficiently long distance, we expect that no calibration is needed and that the sole knowledge of $g_0$ is required which could be obtained through spin-dependent scattering experiments on large samples~\cite{Lo2011}, for example. This observation allows us to propose two different driving modes: mode I, for which we drive the qubit at a frequency $\omega = \bar{\omega}_q = \bar{g}B_0$; and mode II, where $\omega = \omega_0 = g_0B_0$.

In the rest of the section, we study the evolution of the infidelity associated with both driving modes. Let us suppose that we shuttle a qubit back and forth and that its movement is described by a triangle wave, recursively defined as:
\begin{equation} \label{eq:back_and_forth_shuttle}
x(t)= \begin{cases}
  d/2-vt, & \text{if}\ t < d/v \\
  -3d/2+vt, & \text{if}\ d/v \leq t < 2d/v \\
  x(t-2d/v), & \text{if}\ t \geq 2d/v
\end{cases}
\end{equation}
with $d$ representing the distance travelled by the electron for one way and $v$ is its (constant) shuttling speed.  Furthermore, we fix the $g$-factor dispersion $\Delta g = 10^{-3}g_0$ to address cases of higher frequency crowding.

\begin{figure}
    \centering
    \includegraphics[width=\linewidth]{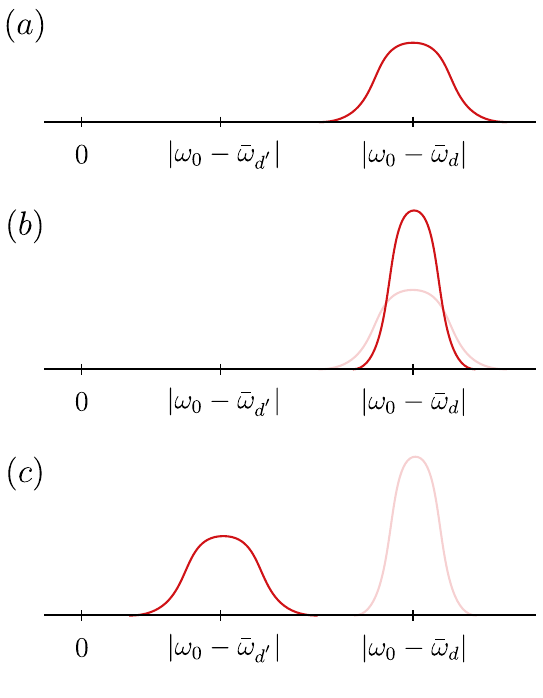}
    \caption{Illustration of the impact of speed and explored distance on the quality of the homogenisation process. The $x$-axis represents the difference between the electron's homogenised frequency $\bar{\omega}_d$ when shuttled for a distance $d$ and the device-averaged one $\omega_0 = g_0B_0$. (a) Wide distribution around the mean frequency representing an imperfect homogenisation. (b) Increasing the speed (\textit{i.e.} the transfer rate) while keeping the explored distance fixed improves the quality of the homogenisation as explained before. This is represented by a more peaked Gaussian centred around $\bar{\omega}_d$, which is reminiscent of the frequency narrowing phenomenon. The faded Gaussian is plotted for comparison. (c) Increasing the shuttled distance while keeping the same speed deteriorates the homogenisation as the electron explores more dots. The shift of the mean frequency manifests the fact that $\bar{\omega}_{d}\rightarrow \omega_0$ as $d\rightarrow\infty$.}
    \label{fig:impact_homogenisation}
\end{figure}

To guide the analysis of the numerical simulations of the infidelity of our protocol, it is important to note that it suffers from two main sources of errors: the frequency homogenisation error and the error arising from off-resonant driving. While the latter is well understood and explained by the Rabi model, we need to describe the impact of $v$ and $d$ on the former. In \cref{fig:impact_homogenisation}, we use Gaussians centred around the mean $g$-factor to symbolise the quality of the shuttling-based frequency homogenisation (in the spirit of frequency narrowing). Considering some distance $d$, and a constant speed $v$ in \cref{fig:impact_homogenisation} (a), the homogenisation is not perfect as illustrated by the wide distribution around the $g$-factor associated with the drive frequency $\omega = \bar{g}_dB_0$. While keeping $d$ fixed, increasing $v$ in \cref{fig:impact_homogenisation} (b) yields a better homogenisation as explained in previous sections. The faded Gaussian, corresponding to a slower speed, is thus less peaked. Keeping $v$ fixed, choosing a distance $d^{'} > d$ in \cref{fig:impact_homogenisation} (c) reduces the quality of the protocol as the electron has to explore more dots. Moreover, as $d\rightarrow\infty$ we have $\bar{g}_{d}\rightarrow g_0$, which explains the shift of the mean frequency. The faded Gaussian, corresponding to a smaller distance, is therefore more peaked but its mean is farther from $g_0$.

\begin{figure}[h!]
    \centering
    \includegraphics[width=\linewidth]{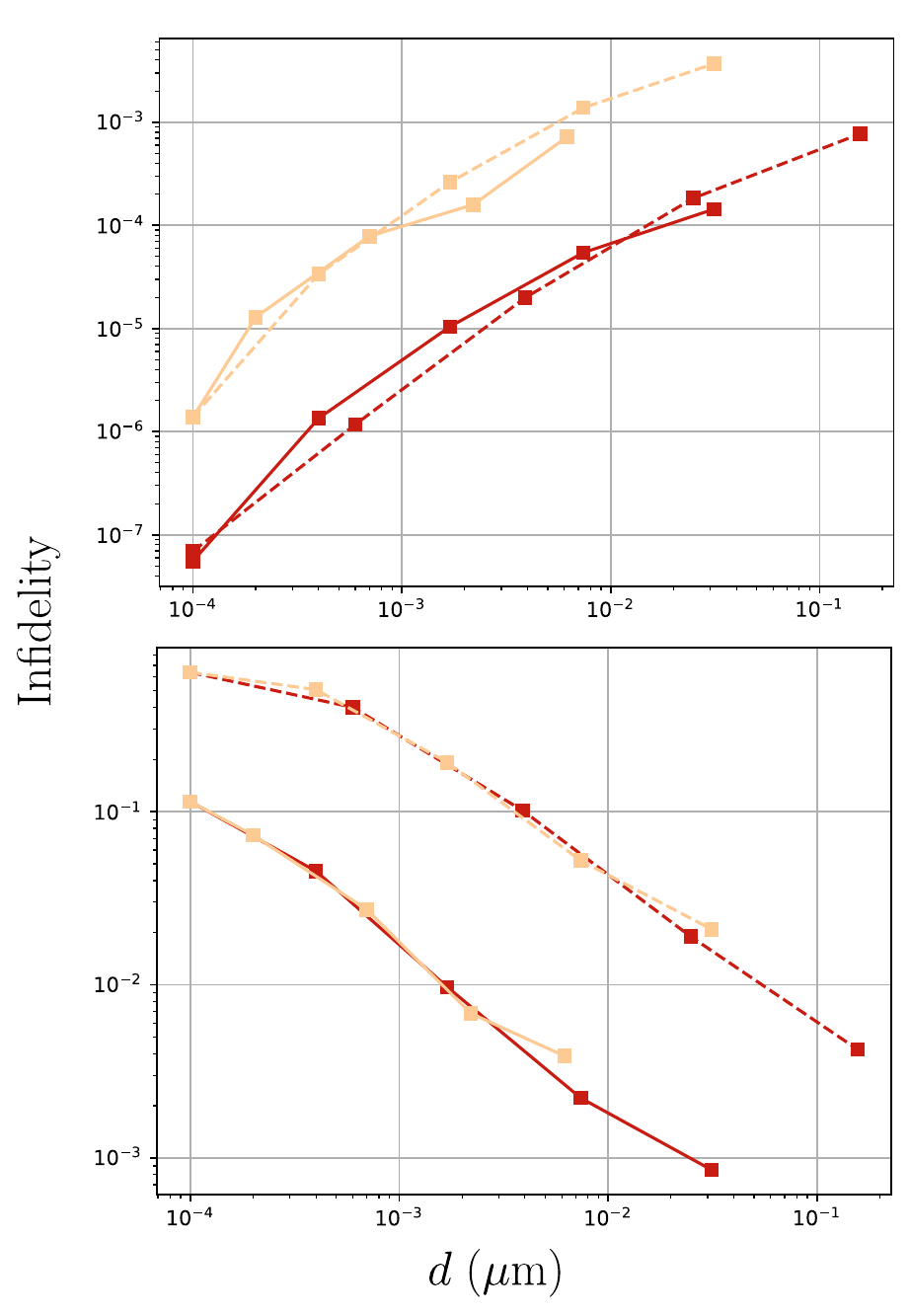}
    \caption{Infidelity of the shuttling-based protocol for modes I (top) and II (bottom) as a function of the distance $d$, the speed $v$ and the drive amplitude $\Omega$. These infidelities are obtained by performing 20,000 Monte-Carlo simulations with random instances of the $g$-factor landscape. The yellow and red lines correspond to $v = 10$ m/s and $v = 50$ m/s, respectively. The solid lines indicates that $\Omega = 5$ MHz, while $\Omega = 1$ MHz for the dashed lines.}
    \label{fig:shuttling_proc_inf}
\end{figure}

We are now ready to study the infidelity of an $X$ gate performed on a qubit as a function of $d$, for different speeds $v$ and drive amplitudes $\Omega$ as plotted in \cref{fig:shuttling_proc_inf} . The distance $d$ ranges from $100$ nm (the interdot distance) to the maximum distance, $L$. As the electron can travel during the gate time $T$, we have $L=vT$. The infidelity measure is the same as for the exchange-coupling-based scheme (further details on this metric and its numerical simulation can be found in \cref{app:numerical_simulations}). Each data point is obtained by averaging 20,000 instances of $g$-factors generated using the method presented in \cref{app:model_g_factor}.

For mode I (top of \cref{fig:shuttling_proc_inf}), in which we drive at the mean frequency $\omega = \bar{g}_dB_0$ ($\bar{g}_d$ being the mean $g$-factor for a distance $d$), the infidelity increases with $d$ for fixed $v$ and $\Omega$. This is characteristic of a degradation of the homogenisation in the presence of a larger number of distinct $g$-factors, as explained before. Then, we notice that the curves can be grouped by their speed $v$, with the infidelities for $v = 50$ m/s (red) approximately an order of magnitude smaller than for $v = 10$ m/s (yellow), in agreement with our predictions. The larger the speed (or transfer rate) $v$, the better the homogenisation. 
Finally, we note that within each group, a change in $\Omega$ (solid versus dashed lines) does not have much impact on the infidelity, at least in the experimentally relevant parameter regime we study. Indeed, when looking at the Rabi model, the value of $\Omega$ does not affect the infidelity when driving on resonance. 

For mode II (bottom of \cref{fig:shuttling_proc_inf}), in which we drive at $\omega = g_0 B_0$, the infidelity decreases with $d$ unlike mode I. Indeed, as $d$ gets larger, $\bar{g}_d$ tends to $g_0$ as showed before, thereby reducing the detuning between the drive and qubit frequency. While we can also group the curves in pairs here, it is now the driving amplitude that discriminates them. Suppose that $\Omega$ is fixed, increasing the speed $v$ only allows for the reduction of the homogenisation error. As we are dominated by the off-resonant driving error here, changing $v$ has a negligible impact. Within the Rabi model, this type of error solely depends on the ratio $|\omega-\omega_q|/\Omega$. The larger the ratio, the higher the fidelity. That is why the infidelity decreases as $\Omega$ increases. 

As our aim is to characterise the quality of the homogenisation, shuttling errors were not included in the previous simulations. If they were, we would need to consider a trade-off between shuttling and homogenisation errors for mode II. Indeed, as the distance increases, the infidelity of our protocol decreases while the cumulative shuttling error would increase. One would need to find a sweet spot balancing the impact of these two error sources. For mode I, increasing $d$ decreases the infidelity, so introducing shuttling error in our calculations would only aggravate the situation. As a reminder, recent experiments in Si/SiGe showed per-increment shuttling fidelities being progressively reduced from $10^{-3}$ to $10^{-4}$ \cite{Yoneda_2021, Seidler_2022, de_smet_shuttling_2024}. As one hop can be implemented in $\sim$10 ns (assuming an interdot distance of 100 nm and a shuttling speed of 10 m/s), the resulting overall shuttling error over one $X$ gate (lasting $1~\mu$s for $\Omega=1$ MHz) reaches at best $10^{-2}$ with current technologies. However, the task of low-noise shuttling is receiving rapidly increasing attention, notably with recent proposals suggesting that $g$-factor disorder might help improve shuttling fidelities rather than hinder them \cite{Bosco_2024}. We will thus assume that shuttling fidelity will keep improving making our proposal easier to realise experimentally.

Additionally, while the results presented in this section are dependent on how the $g$-factor landscape is defined, the homogenisation principle should work for any model. 
However, it is worth noting that this scheme only performs well in the case of low frequency dispersion $\Delta\omega_q$ \textit{i.e.} at low magnetic field $B_0$ and low $g$-factor dispersion $\Delta g$. This is analogous to the observations of \cref{fig:swap_2_qubits} for the case of exchange oscillations, in which the infidelity rapidly degraded with the frequency spread $\omega_q^{12}$. We will confirm this statement for shuttling as well in \Cref{sec:applications}.

Finally, note that we assumed here (and in the rest of the paper) that the drive amplitude $\Omega$ is homogenous over the shuttling path, which can be quite challenging to achieve but not impossible \cite{vahapoglu_single-electron_2021}. We, however, confirmed with additional numerics that our scheme performs the same in the presence of a non-uniform driving field. Namely, the resonant frequency of a shuttled qubit would still be the average $g$-factor over the shuttling path and the resulting infidelity would be identical to the case of uniform $\Omega$. The only distinction resides in the gate time, now defined as $T=\pi/\bar{\Omega}$, where $\bar{\Omega}$ is the average Rabi frequency over the shuttling path.

\subsubsection{Single-qubit gates on more than one electron}

Although we verified here that shuttling could advantageously be used to homogenise $g$-factors in the simplest case of a single qubit, this method can trivially be extended to higher qubit counts. Unlike the previously presented exchange-based scheme, conveyor-belt shuttling \cite{Seidler_2022, xue_qubus_2024} can conveniently be used to synchronously shuttle large groups of non-interacting qubits without increasing resource overhead and without prohibitive reductions in fidelity. Indeed, this mode of shuttling uses only four input voltages to create a moving sine wave, whose minima can hold the shuttled qubits and displace them altogether. Clusters of same-frequency qubits can thus be formed by shuttling qubits around loops, such that all qubits in the loop explore the same $g$-factor landscape.

\subsubsection{Extension to two-qubit gates}

While the focus of this paper is on the use of frequency homogenisation to perform global single-qubit gates, our protocol can be generalised to two-qubit gates as well. With silicon-spin qubits, these can be implemented by turning on the exchange interaction between neighbouring electrons. Depending on additional parameters \textit{e.g.} the existence of a sufficient energy separation between the electrons or the application of a resonant drive, various kinds of gates can be implemented. These include $\sqrt{\text{SWAP}}$, CZ gates and CROT gates \cite{burkard_2023}. We will here focus on the latter as it requires the use of a drive, therefore making its description a natural extension of the single-qubit case.

More precisely, a CROT gate is a rotation around an axis in the $xy$ plane of the spin of a first electron, conditional on the state of a second electron. To implement it, the exchange interaction $J$ must be turned on but kept small compared to the qubits' energy separation. In this parameter regime, the degeneracy between the basis states of the two qubits is lifted, enabling the selective targetting of a specific energy transition with a driving pulse. By setting the drive frequency at the energy splitting of the $\ket{10}\leftrightarrow\ket{11}$ transition, a CROT gate is implemented in a time $\pi/\Omega$ (when $J \ll |\omega_{q,1}-\omega_{q,2}|$). This frequency is given by \cite{noiri_2022, kalra_robust_2014, Russ_2018, Zajac_2018}:
\begin{equation} \label{eq:2q_resonant_freq}
    \omega = \frac{1}{2} \left[\omega_{q,1}+\omega_{q,2} - J + \sqrt{J^2+(\omega_{q,1}-\omega_{q,2})^2}\right]
\end{equation}
where $\omega_{q,i}=g_iB_0$ is the intrinsic frequency of qubit $i\in\{1,2\}$. Note that other energy transitions can also be targeted by tweaking \cref{eq:2q_resonant_freq} accordingly. However, like in the single-qubit-gate case, when many qubits are present in the device, it may prove challenging to selectively target pairs of static qubits due to frequency crowding.

\begin{figure}
    \centering
    \includegraphics[width=\linewidth]{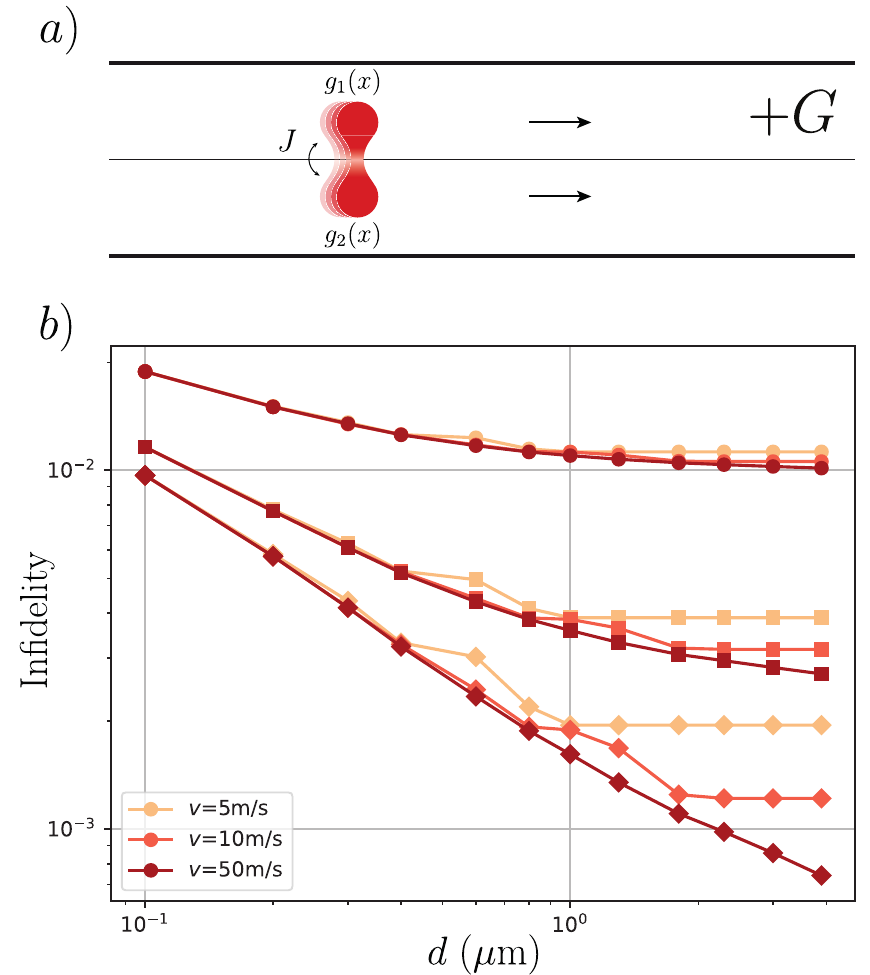}
    \caption{(a) Illustration of the homogenised two-qubit gate. Two exchange-coupled electrons are shuttled along a line in order to homogenise their frequency. A magnetic material is used to effectively shift the $g$-factor of the top row with respect to the bottom one by $G$. (b) Infidelity of a CNOT gate implemented while shuttling the qubits so as to homogenise their respective $g$-factors. Lighter to darker colours correspond to increasing shuttling speeds. Each triplet of curve (from top to bottom: circles, squares, diamonds) is characterised by a distinct frequency gradient $GB_0$ between the shuttling rails, respectively 500 MHz, 1 GHz and 2 GHz.}
    \label{fig:2q_gate}
\end{figure}

Here, we show that frequency crowding can be mitigated by applying our shuttling-based homogenisation scheme in the same fashion as before. We thus consider two electrons evolving on two parallel shuttling tracks characterised by distinct $g$-factor profiles $g_1(x)$ and $g_2(x)$ as illustrated in \cref{fig:2q_gate} (a). A large average frequency difference $GB_0$ is applied between the shuttling tracks \textit{i.e.} by using magnetic materials or a micromagnet~\cite{Kawakami_2014}. Both electrons are synchronously shuttled back and forth on their respective shuttling track over a distance $d$ (see \cref{eq:back_and_forth_shuttle}), with a small exchange interaction $J\ll GB_0$ always turned on between them. The effect of shuttling is to homogenise the respective $g$-factors of the electrons, leading them to acquire the effective frequencies $\bar{\omega}_{q,1}$ and $\bar{\omega}_{q,2}$. In light of the previous section, two driving modes are conceivable: Driving mode I consists in a driven CROT that incorporates the frequency means of the two paths:
\begin{equation}
    \omega_I = \frac{1}{2} \left[\bar{\omega}_{q,1}+\bar{\omega}_{q,2} - \bar{J} + \sqrt{\bar{J}^2+(\bar{\omega}_{q,1}-\bar{\omega}_{q,2})^2}\right]
    \label{eq:crot_freq_mode_1}
\end{equation}
which requires a precise knowledge of the $g$-factor landscape and the average exchange coupling $\bar{J}$. Note that $J$ is being averaged in a similar way to the qubit frequencies. In contrast, in driving mode II, the driven CROT uses the sample average frequency which for one of the paths is shifted by $GB_0$:
\begin{equation}
    \omega_{II} = \frac{1}{2} \left[(2g_0+G)B_0 - J_0 + \sqrt{J_0^2+(GB_0)^2}\right].
    \label{eq:crot_freq_mode_2}
\end{equation}
This is agnostic of the details of the $g$-factor landscapes and uses an exchange coupling $J_0$ that could be set to e.g. $J_0 = J(x = 0)$. Such a choice can be made as the deviations between $\bar{J}$ and $J_0$ can be neglected since $\delta J < J \ll GB_0$. Although less precise, we decide to solely focus here on the more experimentally friendly driving mode \textit{i.e.} mode II.

The results of such an approach are presented in \cref{fig:2q_gate} (b). We observe like in \cref{fig:shuttling_proc_inf} (bottom) that the infidelity decreases with the shuttling distance $d$. Indeed $\omega_{II}$ becomes a better and better estimate of the optimal driving frequency $\omega_I$. Contrary to the single-qubit case however, the infidelity saturates at large $d$ instead of keeping decreasing: this is because even the static and exactly-resonant implementation of the two-qubit gate is imperfect (\textit{e.g} due to crosstalk with other energy transitions, see \cite{noiri_2022, kalra_robust_2014} for details). This source of infidelity can be tamed by increasing $G$ -- since the condition $J \ll GB_0$ would become better enforced -- or by carefully choosing $\Omega$ in order to synchronise the resonant and off-resonant rotations \cite{Russ_2018}.
It is worth noting that increasing $v$ reduces the infidelity too, since it improves the quality of the shutting-induced homogenisation. Finally, note that the scheme we here advocate for only works when electrons are shuttled in separate tracks that are well separated in frequency (large $G$). If $G$ was null or if the electrons were to be placed on the same shuttling track, the resulting operation would be an $XX$ gate.

\section{\label{sec:applications} Applications}
In the previous Section, we showed that the shuttling-based protocol, as opposed to the exchange-based protocol, was more scalable. Indeed, it can be applied to multiple electrons without generating entanglement between them, and it guarantees a high single-qubit gate fidelity while only using a limited number of electrodes. In this Section, we thus focus on shuttling and evaluate its performance for the global control of single-qubit gates in technologically relevant device topologies.

\subsection{Problem statement} \label{subsec:problem_statement}
 
Given the model presented in \cref{sec: preliminary}, we wish to perform a high-fidelity $X$ gate on $n_t$ targets qubits while keeping the other $n_q-n_t$ qubits idle by using a global driving field.

\begin{figure*}
    \centering
    \includegraphics[width=\linewidth]{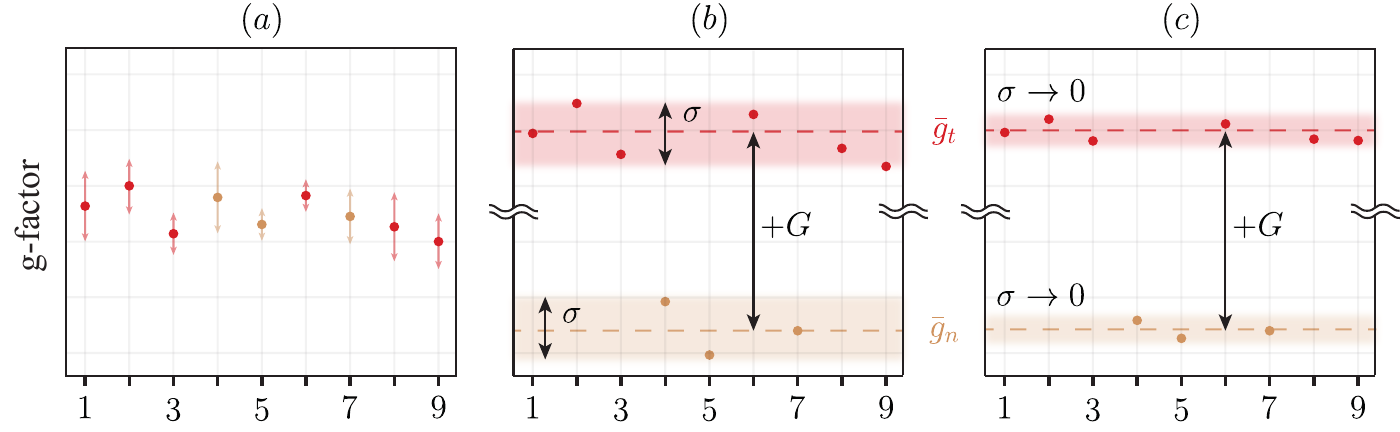}
    \caption{Illustration of the global control workflow. We wish to perform an $X$ gate on the target qubits (red) while leaving the non-target qubits (yellow) idle. (a) Given an initial $g$-factor distribution, we shuttle qubits so as to homogenise their $g$-factors. The arrows describe the $g$-factor landscape explored by each electron and the central dots correspond to the resulting mean values. The quality of this process is in particular dependent on the shuttling speed $v$ (see \cref{fig:impact_homogenisation}). (b) Using micromagnets or superconducting lines, we apply a magnetic field which effectively shifts the $g$-factors of the targets with respect to the non-targets by $G$ as explained in the main text. The target and non-target qubits homogenised $g$-factors are now separated and spread around their means $\bar{g}_t$ and $\bar{g}_n$ (dashed lines). The standard deviation of this distribution is given by $\sigma$ (red and yellow shaded areas). While $\sigma$ could be different for both areas, we suppose that there are equal for simplicity. (c) Our shuttling-based homogenisation allows us to decrease $\sigma$ by increasing the shuttling distance $d$ (see \cref{fig:impact_homogenisation}(c)). In the limit of large $d$, every qubit within the red and yellow shaded areas acquire the same $g$-factor, thereby reducing the spectrum to only one target and one non-target frequencies. Alternatively, $\sigma$ can be negated by further frequency engineering, namely by applying a distinct $G$ to each qubit.}
    \label{fig:workflow}
\end{figure*}

The general workflow we adopt is the following (see \cref{fig:workflow}): we start from an initial distribution of $g$-factors $(g_i)_{1\leq i\leq n_q}$, where we apply both the shuttling-based homogenisation protocol and additional magnetic or electric fields (\textit{e.g.} Stark shift, global magnetic field gradients) to reduce the frequency spectrum to distributions around two $g$-factors, $\bar{g}_t=g_0+G$ and $\bar{g}_n=g_0$, corresponding to the mean target and non-target $g$-factors respectively after homogenisation. There will remain a distribution of target and non-target $g$-factors around $\bar{g}_t$ and $\bar{g}_n$, but we will ensure that the width $\sigma$ of such a distribution is small compared to $G$. The channel frequency spread $\sigma$ is controlled by the shuttling distance $d$: at finite distance $\sigma<\Delta\omega_q$ (frequency homogenisation) while in the infinite distance limit, $\sigma\rightarrow0$. In contrast, $G$ is adjusted via the application of the aforementioned magnetic and/or electric fields. Note that the shuttling speed $v$ has no impact on $\sigma$ (which quantifies the distance between the homogenised $g$-factors and their means $\bar{g}_t$ and $\bar{g}_n$). The speed $v$ only influences the quality of each homogenisation (see \cref{fig:workflow}).

Thereon, we send a driving tone at (almost) resonance with the target qubits: $\omega=\bar{g}_t B_0$. Ideally, when $G$ is large and $\sigma$ is small, the non-targets are left almost idle while each target rotates at a rate $\Omega\approx\bar{g}_t B_1$. In reality, each target rotates around a slightly tilted axis as $\sigma$ is finite. Besides, the non-targets may experience some crosstalk due to finite $G$.
In order to mitigate it, we additionally tune $\Omega$ such that the driving tone induces a $\pi$ rotation on the targets qubit, but a $2\pi$ rotation on the non-targets. This is explained in Appendix \ref{app:crosstalk_sync}.

We will compare our results to those obtained with the binning method, which also allows for the implementation of global single-qubit gates \cite{patomäki2023pipeline, Fayyaz_2023}.
In this case however, it is generally not possible to reduce the spectrum to only two main frequencies like before ($\sigma$ would be too large). Rather, frequency crowding and frequency separation are dealt with by the sole application of electric fields, namely by placing the qubits frequencies into (more than two) same-frequency bins. In this way, the minimum frequency spacing does not decrease with the system size, providing a solution to the frequency crowding issue. The driving pulse must consist of multiple driving tones, separated by a frequency distance equal to the spacing of the bins $2\delta \omega_q$. This relatively low distance can be a source of crosstalk, whose effect is mitigated by tuning the drive strength $\Omega$ like before (see Appendix \ref{app:crosstalk_sync}).

More concretely, we will focus on two architectures: $(i)$ the near-term 2$\times$N quantum dot arrays that can be readily manufactured \cite{Hutin2019,adam_two_by_N_2024} and $(ii)$ the long-term looped pipeline architecture that, while being a two-dimensional architecture, enables qubit connectivity in three dimensions but requires manufacturing advances~\cite{Cai_2023}. Exploring these two architectures allows us to survey the efficacy of our scheme for both the near-term and fault-tolerant eras. Note, however, that 2$\times$N arrays of qubits will also be relevant in the fault-tolerant regime~\cite{adam_two_by_N_2024, siegel_snakes_2025}.

For the study, we set  $g$-factor dispersion $\Delta g$ to $10^{-3}g_0$ or $10^{-2}g_0$, which we will denote as low and high frequency spread regime, respectively. The maximum Stark shift $\delta g$ is set to $\Delta g/10$. Besides, we will use a drive strength $\Omega$ around 5 MHz (taking into account the synchronisation of the target and non-target qubits rotations, as explained in the previous paragraphs and in Appendix \ref{app:crosstalk_sync}). When evaluating the performance of our proposal, we will assume a constant magnetic field $B_0=0.1$ T to minimise frequency spreading and facilitate our homogenisation protocol. In contrast, for the binning method, we will set $B_0=1$ T to maximise the bin separation. As such, we are comparing both schemes in the regime in which they perform best. In addition, we will separate the targets and non-targets by a frequency shift $GB_0=300$ MHz. Assuming that targets and non-targets are on distinct shuttling tracks that are separated by 100 nm (as it will be the case in the following), this translates into a magnetic field gradient of 0.1 mT/nm. This is comfortably below demonstrated gradients induced by micromagnets, around 0.8 mT/nm \cite{KlemtBernhard2023Emoa}.
As for the shuttling distance and speed, we will consider two distinct parameter regimes that will depend on the architecture: $v=10$ m/s and $d=3 \;\mu$m (2$\times$N); or $v=50$ m/s and $d=20 \;\mu$m (looped pipeline). We will comment on the choice of these parameters in \cref{app:param_estimation}. The performance of each technique is evaluated by running 10,000 Monte Carlo simulations with randomly-generated $g$-factor profiles. In each instance, we use the same fidelity measure as before: details on our numerics can be found in \cref{app:numerical_simulations}. Shuttling noise will still be neglected in the present study in order to focus on qubit addressability.

\subsection{\label{subsec:two_by_n} 
Performance on the 2$\times$N array}
\begin{figure*}
    \centering
    \includegraphics[width=\linewidth]{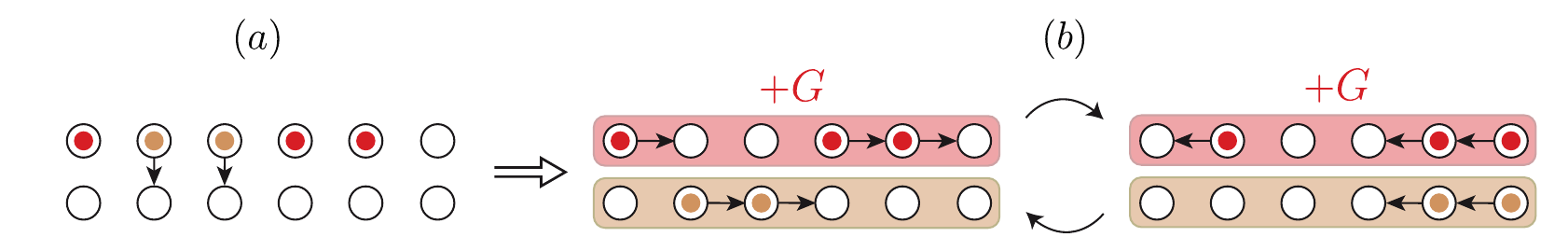}
    \caption{Illustration of the shuttling-based protocol for a 2$\times$N array. Starting with all the qubits on the top row in (a), we move the non-targets to the bottom row to physically separate the two types of qubits. Then, thanks to superconducting lines or micromagnets we apply a large magnetic field gradient between the rows (shaded areas) in (b). This magnetic field gradient can also be seen as shifting the $g$-factors of the electrons on the top row by $G$.
    Finally, the electrons are shuttled back and forth in lines to homogenise their $g$-factors.}
    \label{fig:2xN_workflow}
\end{figure*}

The first architecture we consider corresponds to a 2$\times$N array of quantum dots. Despite its simplicity, it has been shown that error correction can be reliably implemented in this architecture \cite{adam_two_by_N_2024}. However, here we are concerned with a more restricted version of the architecture in which we require that the second row of the 2$\times$N array be empty as shown in \cref{fig:2xN_workflow}. In this section, we will assume a varying number of target qubits $n_t$ and set the number of non-target qubits to $n_n=n_t-2$ as an example.

As explained in the previous section, our first goal is to simplify the frequency spectrum, namely to only retain one main target $g$-factor $\bar{g}_t$ and one main non-target $g$-factor $\bar{g}_n$. To do so, we first spatially separate the targets from the non-targets by transferring the non-target qubits to the bottom row (\cref{fig:2xN_workflow}(a)). In addition to this, we apply a magnetic field gradient between the rows, \textit{e.g.} by placing micromagnet or a superconducting wire parallel to the array. This generates the desired frequency separation $GB_0$. At this stage however, the target and non-target $g$-factors are scattered around their mean values (respectively $\bar{g}_n = g_0$ and $\bar{g}_t = g_0+G$). To lower the standard deviation $\sigma$ of these dispersions, we shuttle both rows of qubits back and forth (\cref{fig:2xN_workflow}(b)). This results in reducing $\sigma$ by homogenising the $g$-factor landscape as illustrated in \cref{fig:workflow}. Target qubits can now be driven close to resonance by sending a driving pulse at $\omega=(g_0+G)B_0$. This is reminiscent of the driving mode II we explored in \cref{subsec:shuttling}, where qubits are not driven at exact resonance, but rather at a known frequency that is agnostic of the details of the $g$-factor landscape ($\omega$ only depends on the device average $g$-factor $g_0$ and the set magnetic field gradient $G B_0$). As such, the infidelity is arising from two main contributions: the slightly off-resonant driving of the target qubits and the crosstalk between targets and non-targets. Note that in the present scheme, it is not feasible to use the driving mode I (where each target qubit is driven at exact resonance). Indeed, the leftmost and rightmost qubits of the top row do \textit{not} experience the same $g$-factor landscape, yet their resulting effective $g$-factors are both too close to $\bar{g}_t$ to differentiate them and drive them with two distinct driving tones.

\begin{figure}
    \centering
    \includegraphics[width=\linewidth]{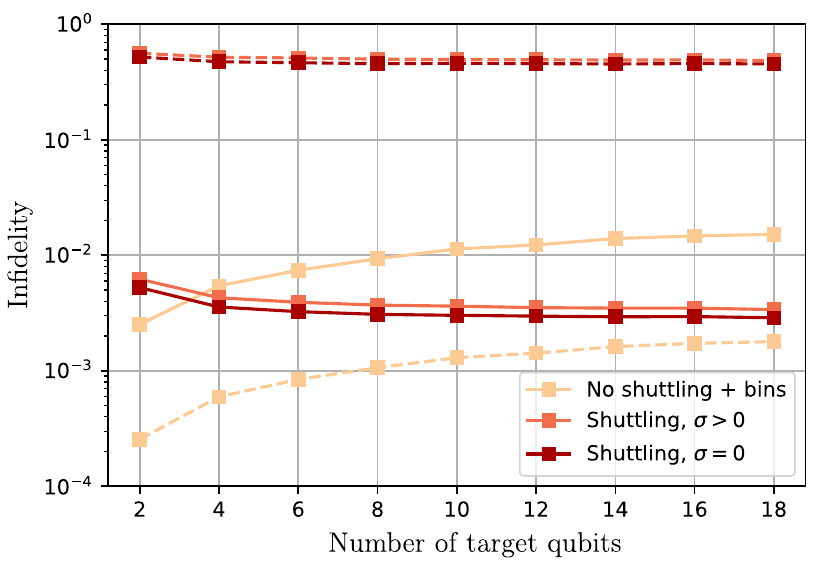}
    \caption{Single-qubit gate infidelity for the 2$\times$N architecture in the presence of $n_t$ target qubits and $n_n=n_t-2$ non-target qubits. The solid and dashed lines respectively correspond to a $g$-factor dispersion $\Delta g=10^{-3}g_0$ and $10^{-2}g_0$. The targets and non-targets are separated by a magnetic field gradient $GB_0$. We consider three different strategies for the gate implementation: binning the frequencies without shuttling the qubits (yellow); shuttling the target and non-target qubits so as to reduce the spreading $\sigma$ of their frequencies (orange); and the same as the latter with additional frequency engineering in order to negate any frequency spreading among the target qubits \textit{i.e.} setting $\sigma=0$ (brown). In the first case, the drive contains multiple tones at resonance with each bin. In the second and third cases however, we send a single tone either approximately or exactly at resonance with the target qubits (owing to $\sigma$ being finite or not).}
    \label{fig:infid_2xN}
\end{figure}

In \cref{fig:infid_2xN}, we compare this protocol with the binning method, as explained in the previous section. We use a shuttling speed $v=10$ m/s and a distance $d=3\;\mu$m \textit{i.e.} spanning 30 quantum dots. This scenario is realistic for near term devices \cite{de_smet_shuttling_2024}. We find that in the case of low frequency spread (solid lines) our shuttling protocol (orange) quickly outperforms the binning scheme (yellow). More specifically, it yields a per-qubit infidelity around $0.3\%$ \textit{i.e.} under the surface code threshold, which the binning scheme fails to provide. We however observe an opposite behaviour in the case of high frequency spread (dashed lines), where the binning scheme yields a desirable sub-threshold single-qubit infidelity, far below the one obtained with shuttling. This is because the homogenisation of the $g$-factors becomes more demanding at large $\Delta g$, requiring unfeasible shuttling speeds or drive strengths to compensate for the wider dispersion when shuttling. In contrast, the binning scheme performs well in high frequency spread, due to a larger interbin spacing arising from a larger Stark shift $\delta g=\Delta g/10$. Note, however, that contrary to our shuttling scheme, the binning method requires a perfect knowledge of the $g$-factor landscape and a reliable tunability via Stark shift. While not achievable in practice, the optimal $\sigma=0$ case is also plotted (darker lines): one can see that for $\sigma>0$ \textit{i.e.} for a realistic parameter regime, we closely approach the performance of this optimal case.

An additional interesting feature one can note is that the average single-qubit infidelity as defined in \cref{eq:fidelity} decreases when loading more qubits into the device with our shuttling scheme. In contrast, it grows with the binning scheme. We can explain this by noting that in the first case we are not increasing the number of driving tones with the number of qubits, while in the second case, a new driving tone should be added every time a new bin is created. This leads to an increase of the crosstalk between target qubits. At sufficiently large numbers of qubits, both infidelities become stationary however, meaning that both schemes can address the frequency crowding issue, albeit in different parameter regimes.

\subsection{\label{sec:3-looped} Performance on the looped pipeline architecture}

\begin{figure*}
    \centering
    \includegraphics[width=\linewidth]{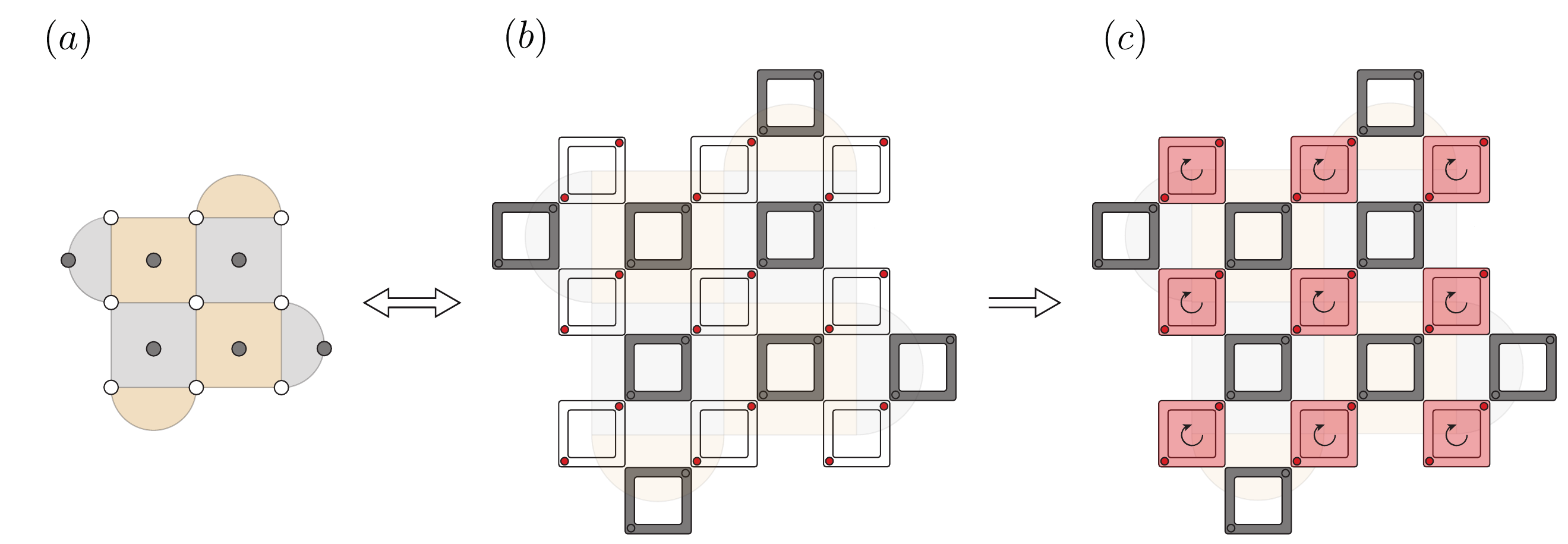}
    \caption{(a) Schematic representation of the $3\times3$ surface code with data and ancilla qubits represented in white and grey respectively. (b) Implementation of the $3\times3$ surface code within the looped-pipeline architecture with white and grey loops representing data and ancilla loops respectively. Note that $n$ electrons ($n=2$ here) can populate each loop in order to encode more than one surface code patch as explained in the main text. Red circles represent data qubits that we wish to target while empty ones represent ancilla qubits that we can ignore here as explained in \cref{app:stab_cycle}.
    (c) Implementation of single-qubit gates using our shuttling-based homogenisation protocol. During each stabiliser cycle, we wish to perform single-qubit gates on the data loops. By shuttling the electrons around loops we homogenise their $g$-factors and reduce the number of targetted distinct frequencies from $9n$ (with $n$ being the number of electrons per loop) to 9 (number of loops).
    Note that, one could add additional magnetic materials to the red loops such that every target obtains the same $g$-factor ($\sigma = 0$), which yields an even higher fidelity as seen in \cref{fig:infid_lpip}
    }
    \label{fig:Lpip_workflow}
\end{figure*}

The second architecture is the looped pipeline, a two-dimensional grid of interacting shuttling loops enables the encoding of 3D error-correcting structures within a strictly 2D device~\cite{Cai_2023}. These features are achieved by placing alternating loops of data and ancilla shuttling qubits that meet at interaction points to perform two-qubit gates; see Fig~\ref{fig:Lpip_workflow}. Each loop can fit multiple electrons, creating out-of-plane layers of error correcting codes ($z$ axis) that are effectively stacked together in physical space (see Fig. 4 of \cite{Cai_2023}).

In this Section, we will focus on the realisation of stabiliser cycles which represent the most frequent operation on a fault-tolerant quantum computer. Throughout their normal execution, data qubits often need to be collectively addressed by single-qubit Hadamard gates while leaving ancilla qubits idle~\cite{Veldhorst2017}.
However, as we show in \cref{app:stab_cycle}, one can implement a stabiliser circuit where the ancilla is initialised after (measured before) the global single-qubit gates on the data qubits. For this reason, the impact of the global gate on the ancillas can be ignored, thus we can set $n_n = 0$.

Thanks to the above feature, we have more flexibility in how we engineer the qubit addressing and propose two scenarios. In the simplest one, we apply no magnetic field gradient, just the global Zeeman splitting field $B_0$. As qubits go around in loops, they acquire an effective frequency equal to the mean frequency of their respective shuttling track. Assuming a perfect loop-wise $g$-factor homogenisation, the number of frequencies in the spectrum is given by the number of loops rather than the total number of qubits. Given the use of a single global drive frequency, the infidelity in this scenario is caused by the slightly off-resonant driving of all the targets ($\sigma>0$).

In the more involved scenario, we propose applying local magnetic fields at the loop level to cancel out the $g$-factor dispersion, \textit{i.e.} $\sigma = 0$, leaving the frequency spectrum with a single frequency only. However, this requires knowledge of all mean $g$-factors $\bar{g}_i$ and the capability to apply loop-specific magnetic fields to compensate their deviations. One could envision using the tunable Oersted fields produced by superconducting coils~\cite{li_crossbar_2018} or in-situ tunable magnetic materials~\cite{Lee2018}. This scheme represents a higher experimental challenge, but yields significant reductions in infidelity, due to the total suppression of the dispersion $\sigma$ of homogenised $g$-factors around $\bar{g}_t$ (see \cref{fig:workflow}). This means that all target qubits can be driven at exact resonance. The only remaining source of infidelity in this case is thus any residual imperfect homogenisation of the $g$-factors.

\begin{figure}
    \centering
    \includegraphics[width=\linewidth]{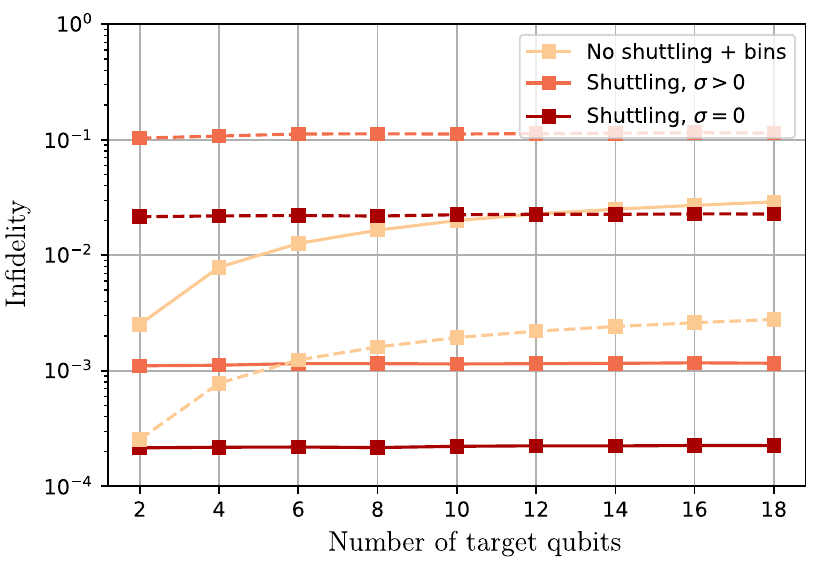}
    \caption{Single-qubit gate infidelity for the looped-pipeline architecture in the presence of $n_t$ target qubits. The solid and dashed lines respectively correspond to a $g$-factor dispersion $\Delta g=10^{-3}g_0$ and $10^{-2}g_0$. We compare the three strategies previously defined in \cref{fig:infid_2xN}.
    }
    \label{fig:infid_lpip}
\end{figure}

We present the performance of these two schemes in \cref{fig:infid_lpip}, along with the binning scheme for comparison. In this case, we set a shuttling distance $d=20\mu$m and a shuttling speed $v=50$ m/s. 
Although the binning scheme is not affected by the choice of $d$ and $v$, we observe a general reduction in the infidelity of our shuttling scheme compared to \cref{fig:infid_2xN}. The main observations remain the same however: shuttling proves advantageous in the low-frequency-spread regime, while binning yields promising infidelities for high frequency spread. The main difference compared to \cref{fig:infid_2xN} is that the $\sigma=0$ data (brown lines) outperforms the $\sigma>0$ data (orange lines) by almost an order of magnitude, potentially justifying the use of this more experimentally-demanding strategy. The reason why it proved less advantageous in the 2$\times$N case was that the homogenisation error was saturating the infidelity at low $v$. 
Moreover, the rightmost data points correspond to two stacked $3\times 3$ rotated surface codes (two qubits per loop), for which we observe comfortable sub-threshold infidelities for the shuttling or binning schemes, respectively in the low or high frequency spread regimes. Since at this point all infidelities are stationary with the number of qubits, we expect this observation to hold for any code size and number of stacked surface codes.

\section{\label{sec:conclusion} Discussion and outlook}

In this work, we consider the pressing issue of individual spin qubit control as spin qubit quantum processors are scaled up. When qubits are encoded with a single spin trapped in a quantum dot, single-qubit gates are implemented via the use of magnetic fields, which are hard to localise at the qubit scale. Rather, existing research advocated for the use of a global field with multiple driving tones at resonance with the target qubit frequencies, which are variable in essence, due to, for example, interface roughness. This generally induces frequency crowding thus unwanted crosstalk that needs to be addressed. In this piece of research, we design a framework based on spin shuttling to globally and selectively control silicon spin qubits with high fidelity. 

More specifically, we demonstrate that spin motion can be utilised to homogenise the $g$-factors. When the motion is based on exchange oscillations between a pair of qubits, both spin $g$-factors become equal to the mean of the original $g$-factors. When the motion is implemented via spin qubit shuttling, all spins travelling around a looped shuttling track obtain the same new effective $g$-factor equal to the loop mean $g$-factor. These examples make it clear that the above homogenisation method can be utilised to reduce the number of frequencies in the spectrum. In the present paper, we put this observation to use for the implementation of single-qubit gates. While frequency narrowing had already been identified in the past \cite{Langrock_2023}, here we explore a new paradigm where it becomes an beneficial ingredient for the implementation of quantum gates at scale.

The main result of this paper is that when the spins are transferred between multiple locations, a single-qubit gate can simply be implemented by sending a driving tone at resonance with the mean frequency felt by the spin. The quality of the gate is dependent on the transfer rate (exchange coupling, shuttling speed): highest fidelities are achieved when the transfer rate is large compared to the Rabi frequency and the qubits frequency dispersion. Via numerical simulations, we quantify the amount of error generated by such a protocol and obtain promising estimates in the case of transferring a single spin via shuttling. In particular, we demonstrate that when the shuttling distance is large enough, the qubit $g$-factor dispersion rapidly decreases, meaning that the sole knowledge of the device average $g$-factor $g_0$ suffices to set the driving frequency. This alleviates the experimental requirements for a spin-qubit-based quantum computer, by removing the need for constant monitoring and calibration of the detailed $g$-factor landscape. These promising observations led us to extend our analysis first to two-qubit gates, for which we still showed the suitability of our protocol. Then we aimed to tackle realistic experimental setups: a 2$\times$N array of qubits and the looped pipeline architecture \cite{Cai_2023}. Even in these more complex contexts, we showed orders of magnitude improvements of the fidelity of single-qubit gates on selected subsets of qubits compared to previous state-of-the-art techniques.

As future work, it would be valuable to generalise our protocol to more general types of qubits and materials. Here, we focused on electron spins in silicon, where the $g$-tensor anisotropy is small. This means that we were able to only consider scalar $g$-factors. More generally, we believe that effects of stronger spin-orbit coupling can be calibrated away as long as they are known \emph{a priori}. 

Additionally, we envision extending our study to more kinds of architectures. One example is the novel Quantum Snakes on a Plane \cite{siegel_snakes_2025}, where logical qubits consist of linearised objects than can be shuttled around large loops made of 2$\times$N filaments. We believe that our protocol could greatly facilitate the implementation of single- and two-qubit cases in this paradigm. More generally, as shuttling becomes an enabling feature of fault-tolerant architectures \cite{adam_two_by_N_2024, siegel_snakes_2025, Cai_2023, boter_spiderweb_2022, kunne_spinbus_2024, otxoa_spinhex_2025, li_crossbar_2018}, the ability to perform gates while moving qubits could greatly reduce the run time of algorithms.

Another direction we wish to explore is the inclusion of various physical and experimental limitations. First, this encompasses off-resonant driving, \textit{e.g.} stemming from the use of finite-bandwidth driving tones. Second, non-adiabatic effects could arise from the abrupt switching-on and -off of the driving pulse and exchange coupling. One could further evaluate if these effects can be tamed via known pulse shaping techniques. Finally, noise would inevitably be present in any experimental chip, \textit{e.g.} charge noise or valley excitations (when shuttling). It would be valuable to run such simulations, which we overlooked for now in order to solely tackle the qubit addressability problem. These would further inform the user about the trade-offs between the use of our schemes and increased noise when qubits are swapped or shuttled. 

\section*{Acknowledgments}
The authors thank Simon Benjamin, Guido Burkard and Andrew Fisher for helpful discussions. The numerical modelling involved in this study made use of the University of Oxford Advanced Research Computing (ARC) facility~\cite{oxford_arc}
and specifically the facilities made available from the EPSRC QCS Hub grant (agreement No. EP/T001062/1). M.F.G.Z.~acknowledges support from the UKRI Future Leaders Fellowship (MR/V023284/1).

\appendix

\section{\label{app:model_g_factor} Modelling the $g$-factor}

The $g$-factor of an electron trapped in a quantum dot in silicon is related to the local spin-orbit fields caused by broken symmetries at the interface where dots form and can be linked to surface roughness, and is thus dependent on the precise position of the electron. As there exists no definitive microscopic description for $g$-factor variations \cite{Langrock_2023, adam_two_by_N_2024}, we decide to model it with a 1D Ornstein-Uhlenbeck (OU) $g_{int}(x)$ process. The restriction to 1D is motivated by the fact that we will be concerned by moving electrons along linear shuttling tracks. We chose this process as it allows for the modelling of a continuous random process with a certain correlation length, yet remaining relatively easy to simulate. This model should allow us to capture meaningful insights on the impact of the interface on our protocol \cite{mokeev_decoherence_2024}.

The OU process is characterised by its mean $g_0$, its standard deviation $\Delta g$ and its correlation length $\lambda$. In silicon, $g_0$ approaches 2 and Ref. \cite{ferdous_interface_2018} estimated that $\Delta g$ lies between $10^{-3}g_0$ and $10^{-2}g_0$. We will study both ends of this range in the paper. Here, $\lambda$ represents the correlation length of the interface roughness which largely dictates the values of the $g$-factors. We will set $\lambda = 20$ nm. To be more precise, $g_{int}(x)$ is given as the solution of the following stochastic differential equation,
\begin{align}
    dg_{int}(x) = \frac{1}{\lambda}(g_0-g_{int}(x))dx+\sqrt{\frac{2dx}{\lambda}}\Delta g dw(x) 
    \label{eq:g_ou_process}
\end{align}
where $w$ represents a Wiener process. 

\cref{eq:g_ou_process} corresponds to the $g$-factor at each position $x$, but does not yet describe the $g$-factor $g_i$ (appearing in \cref{eq:hamiltonian}) of an electron at the position $x_i$. One must indeed additionally take into account the spatial delocalisation of the electron around its central position $x_i$. To do so, the electron $g$-factor should rather be defined as the average of $g_{int}(x)$ weighted by the modulus squared of the wavefunction.

Consider a single quantum dot centred at $x=x_i$, the charge wavefunction of an electron trapped in this dot can be modelled as a Gaussian centred on $x_i$ such that, 
\begin{align}
    |\psi(x,x_i)|^2 = \frac{1}{\mathcal{N}}\exp(-\frac{(x-x_i)^2}{2\lambda_d^2}),
\end{align}
with $\lambda_d = 7$ nm nm which leads to a realistic size of the electron's wavefunction of approximately $40$ nm (the electron is in the region $\pm3\lambda_d$ with a probability of $99\%$) and $\mathcal{N} = \int_{\mathbb{R}}|\psi(x,x_i)|^2dx$ is a normalisation constant. For simplicity, we only consider the wavefunction in 1D as we do not expect it to be heavily modified in the other directions whilst the electron is being shuttled. Note however that our scheme works for any model of $g$.
The averaged $g$-factor $g_i$ is thus defined by,
\begin{align}
    g_i = g(x_i) \coloneqq \int_{\mathbb{R}} |\psi(x,x_i)|^2g_{int}(x)dx.
    \label{eq:avg_g_factor}
\end{align}

\begin{figure}[h!]
    \centering
    \includegraphics[width=\linewidth]{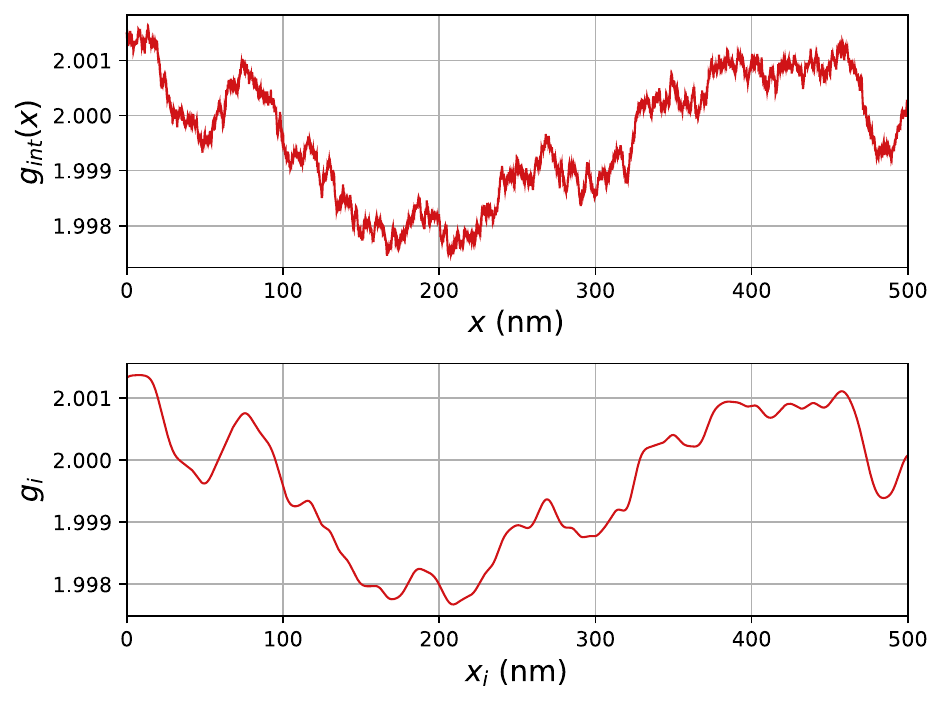}
    \caption{Evolution of one instance of the random $g$-factor $g_{int}(x)$ and the corresponding averaged $g_i$. $g_{int}(x)$ is given by a 1D Ornstein-Uhlenbeck process of mean $g_0 = 2$ and standard deviation $\Delta g = 10^{-3}g_0$. The averaged $g$-factor is obtained by averaging the position-dependent $g$-factor $g_{int}(x)$ with respect to the electron's wavefunction $\psi(x, x_i)$ as in \cref{eq:avg_g_factor}. }
    \label{fig:g_factors}
\end{figure}

In \cref{fig:g_factors}, we plot one instance of $g_{int}(x)$ and the corresponding evolution of $g_i$.

Note that in the main text we drop the subscript $i$ and use $x$ to denote the position of the centre of the electron's wavefunction when the context is clear.

\section{\label{app:g_tensor} Impact of using the g-tensor}
In \cref{eq:hamiltonian}, we use the following Hamiltonian to describe the impact of a magnetic field on the spin of an electron,
\begin{align}
    H_{mag} = g \vec{S} . \vec{B},
\end{align}
where $g$ and $\vec{S}$ represent the $g$-factor and spin of the electron respectively, and $\vec{B}$ the applied global magnetic field. 
However, as explained in the main text, the most precise way of describing this interaction requires the use of the full $g$-tensor $\mathbb{G}$. In this case, the Hamiltonian is now given by,
\begin{align}
    H_{mag} = \vec{S}^{T}\mathbb{G}\vec{B}.
\end{align}
Simply put, the effect of $\mathbb{G}$ is to change the axes of the applied magnetic field $\vec{B}$, ultimately leading to the modification of the single-qubit gates axes. As this change is position-dependent, this could then hinder our shuttling-based homogenisation protocol presented in \cref{subsec:shuttling}.

Considering a driving field along a single axis and 
the matrix $\mathbb{G}$ associated to Si/SiO$_2$ heterostructures, we show here that using the $g$-tensor has minimal impact on the axes of single-qubit rotations. We even show that this impact could be negated, provided perfect knowledge of $\mathbb{G}$.
Suppose that $\vec{B} = \vec{B}_\text{osc} + \vec{B}_\text{stat}$
with $B_\text{osc} = (B_1\cos(\omega t+\phi), 0, 0)^T$ and $B_\text{stat} = (B_x, B_y, B_0)^T$ its oscillatory and static parts. While the $z$-component of the Zeeman splitting field is dominant, smaller $x$- and $y$-components can arise from our use of micromagnets or superconducting lines. Moreover, we will use a matrix $\mathbb{G}$ obtained from atomistic simulations performed by Cifuentes et al. in \cite{cifuentes_bounds_2024},
\begin{align}
    \mathbb{G} = \begin{pmatrix}
        g_0 + \alpha & \beta & g_{13} \\
        \beta & g_0 + \alpha & g_{23} \\
        0 & 0 & g_{33} 
    \end{pmatrix},
\end{align}
where $g_0 \approx 1.994$, $\alpha \approx -10^{-3}$, $\beta \ \approx \pm 10^{-2}$, $g_{13} \approx \pm 10^{-3}$, $g_{23} \approx \pm 10^{-3}$ and $g_{33} \approx 2.002$ are typical values of the different parameters for Si/SiO$_2$. 
We can now expand $H_{mag}$ as follows,
\begin{align}
    H_{mag} &= \frac{1}{2}\Big( (g_0 + \alpha)(B_1\cos(\omega t+ \phi) +B_x) \nonumber \\
    &+ \beta B_y + g_{13}B_0\Big)\sigma_x \nonumber \\
    &+ \frac{1}{2}\Big( \beta (B_1\cos(\omega t+ \phi) + B_x)\nonumber \\
    &+(g_0+\alpha)B_y + g_{23}B_0\Big)\sigma_y \nonumber \\
    &+ \frac{1}{2}g_{33}B_0\sigma_z,
\end{align}
where $\sigma_x, \sigma_y$ and $\sigma_z$ are Pauli operators acting on the spin degree of freedom.

We now consider $H_{mag}$ in a frame rotating at the drive frequency $\omega$ by performing the following transformation,
\begin{align}
    H_{mag}(t) &\rightarrow H_{mag}^{'}(t) = U(t)H_{mag}(t)U^{\dagger}(t) +i\frac{dU}{dt}(t)U^{\dagger}(t),
\end{align}
with
\begin{align}
    U(t) = \exp(i\frac{\omega}{2}\sigma_z t).
\end{align}
Using the rotating-wave approximation we get, 
\begin{align}
    H_{mag}^{'} &= \frac{1}{4}g_0\tilde{B}_1\cos(\theta +\phi) \sigma_x \nonumber \\
    &+ \frac{1}{4}g_0\tilde{B}_1\sin(\theta +\phi) \sigma_y\nonumber \\
    &+ \frac{1}{2}(g_{33}B_0-\omega)\sigma_z,
\end{align}
where $\cos(\theta) = (1+(\beta/(g_0+\alpha))^2)^{-1/2}$ and $\tilde{B}_1 = \sqrt{(1+\alpha/g_0)^2+(\beta/g_0)^2}B_1$. 

More precisely, supposing that $\phi = 0$, instead of obtaining a rotation along the $x$-axis with Rabi frequency $g_0B_1$ (as expected from the Rabi model), the use of the $g$-tensor leads to a rotation along the axis $(\cos(\theta), \sin(\theta), 0)^T$ with a slightly modified frequency $g_0\tilde{B}_1$. 
Using the values extracted from \cite{cifuentes_bounds_2024}, we find that $\sin(\theta) \approx O(10^{-3})$ and $|B_1-\tilde{B}_1|/B_1 \approx O(10^{-4})$, which has a negligible impact on the gate fidelity. 
Finally, we can approximate $H^{'}_{mag}$ by,
\begin{align}
    H^{'}_{mag} \approx \frac{1}{4}g_0B_1\sigma_x + \frac{1}{2}(g_{33}B_0-\omega)\sigma_z.
\end{align}
which validates our assumption that the oscillatory and static fields are respectively applied along the $x$ and $z$ axes.
For commodity we will additionally assume that $g_0\approx g_{33}$, which allows us to consider a scalar $g$-factor.

Alternatively, if we have perfect knowledge of $\mathbb{G}$, we can control the rotation axis and obtain a perfect rotation around the $x$-axis by setting $\phi = -\theta$. The rotation frequency $\tilde{B}_1$ would also be fully characterised.

\section{Numerical simulations} \label{app:numerical_simulations}

To quantify the success of each technique we present in this paper, we measure the fidelity of the implemented operation via the matrix norm:
\begin{equation} \label{eq:exact_fidelity}
    \mathcal{F} = \frac{1}{4}\left|\tr(U_{\text{target}}^\dagger \tilde{U})\right|^{2/n_q}
\end{equation}
where $U_{\text{target}} = X^{\otimes n_t} \otimes I^{\otimes n_q-n_t}$ is the target unitary and $\tilde{U}$ is a slightly modified version of the time evolution operator $U$ as explained in \cref{eq:modified_u_z_rot}. Note that this expression is an average single-qubit fidelity over the $n_q$ qubits, which is why an $n_q$-th root is taken. As $H$ is time-dependent, we compute $U$ by discretisation of the time, with time step $dt=T/N_{\text{steps}}$, where $T=\pi/\Omega$ is the total gate time of an $X$ gate with drive strength $\Omega$. $N_{\text{steps}}$ is set to $10,000$, such that the simulations converge.

When the qubits are non-interacting and as this time discretisation induces a computational overhead, we simplify the calculations by further decomposing the fidelity $\mathcal{F}$ into a product of single-qubit fidelities. These single-qubit fidelities are then slightly modified, by noting that \textit{known but unwanted} $Z$-rotations can easily be cancelled out by a change of rotating frame (or equivalently via the implementation of a virtual $Z$ gate) \cite{McKay_2017}. Therefore, when evaluating the success of a single-qubit gate, one should remove the $Z$ components of the implemented unitary $U$. 

A general representation of any $SU(2)$ gate is:
\begin{align}
    U &= \begin{pmatrix}
        \mathrm{cos}(\theta/2) & -i\mathrm{e}^{i\lambda}\mathrm{sin}(\theta/2) \\
        -i\mathrm{e}^{i\phi}\mathrm{sin}(\theta/2) & \mathrm{e}^{i(\lambda+\phi)}\mathrm{cos}(\theta/2) \\
    \end{pmatrix} \\
       &= Z_\phi X_\theta Z_\lambda
\end{align}
with $P_\alpha = \mathrm{e}^{-i\alpha P/2}$ and $X$ and $Z$ the Pauli operators. Removing the $Z$-rotations from $U$ thus amounts to setting $\phi=\lambda=0$. As $\theta\in[0,\pi[$, this is equivalent to considering the modified time-evolution operator
\begin{equation}
    \tilde{U} = \begin{pmatrix}
      |u_{0,0}| & -i|u_{0,1}|\\ 
      -i|u_{1,0}| & |u_{1,1}|
    \end{pmatrix}
    \label{eq:modified_u_z_rot}
\end{equation}
with $u_{i,j}$ the coefficients of the matrix $U$.

The case of two-qubit gates (for the exchange-coupling scheme) is more complex and was not implemented here, which means that we will not attempt in this case to remove unwanted $Z$ rotations.

Finally, note that we will only study unitary evolutions in this paper, thus not simulating noise processes such as charge noise or shuttling noise. If we were to include them in our simulations, the resulting infidelities would only saturate at a level given by the strength of the dominant noise source.

\section{\label{app:schmidt} Residual entanglement in the exchange-based homogenisation}

In the exchange-based homogenisation scheme, we noticed that the achieved fidelities were somehow limited. In an attempt to better understand the unitary operation $U$ resulting in the simultaneous driving and swapping, we decided to extend our study to its Schmidt decomposition. Mathematically, any vector in the tensor product of two vector spaces can be expressed as a sum of tensor products. Applied to $U$, this reads:
\begin{equation}
    U = \sum_{i=1}^m \alpha_i V_i \otimes W_i
\end{equation}
where for all $i\in\llbracket 1,m\rrbracket$, $V_i$ and $W_i$ are single-qubit unitary operations on the first and second qubit respectively, and $\alpha_i \in \mathbb{C}$. This decomposition is thus a very convenient way to deduce if a two-qubit unitary is a product of two single-qubit unitaries, in which case only one $\alpha_i$ should be non-zero. In our setup, we verified that the ideal case of instantaneous swapping led to a single non-vanishing $\alpha_i$ with the expected unitaries: two single-qubit rotations around the $x$-axis at frequency $\Omega$. We then aimed to understand if beyond this ideal case, the resulting unitary $U$ could be written as the product of two single-qubit rotations, potentially around slightly off axes or with slightly different Rabi frequencies. We however found that reducing the ratios $J/\Omega$ or $\Omega/\omega_q^{12}$ inevitably led, even to first order, to more than one non-vanishing $\alpha_i$. Finite values of such ratios thus necessarily, but also quite expectedly, increase the entanglement power of the operation. This results in an unwanted mixing of the single-qubit states, that seems to be the first source of infidelity: the only solution against this thus is to operate with higher values of the above ratios. The easiest way to do so is to increase the exchange coupling $J$. However, it might be experimentally challenging to operate with large exchange couplings.

\section{\label{app:tunnnel_coupling} Tunnel-coupling-based homogenisation}

\subsection{\label{app:hamiltonian_tc} Hamiltonian}

In the main text, we noticed that the achievable values of the $J$-coupling were a limiting factor in the fidelities obtained with the exchange-based homogenisation scheme. We then noted that tunnel oscillations could be implemented at a higher rate than exchange oscillations, at the cost of placing a single electron per DQD, rather than two. In this case, one needs to take into account the charge degree of freedom which is not explicitly apparent in \cref{eq:hamiltonian}. The homogenisation induced by the usage of the tunnel coupling is based on the fact that the left and right dots have different frequencies $\omega_{q,1}$ and $\omega_{q,2}$ respectively and that the charge of the electron will oscillate between the two when the barrier is lowered. 

The Hamiltonian in the laboratory frame of reference is given by:
\begin{align}
    H_{single} &= \frac{\omega_{q,1}}{2}\ketbra{\text{L}}{\text{L}}\otimes\sigma_z + \frac{\omega_{q,2}}{2}\ketbra{\text{R}}{\text{R}}\otimes\sigma_z  \nonumber \\ &+ \varepsilon \tau_z \otimes I + t_c \tau_x\otimes I \nonumber \\&+ \Omega \cos(\omega t + \phi) I \otimes \sigma_x,
    \label{eq:single_electron_hamil}
\end{align}
where $\varepsilon$ is the DQD detuning, $\tau_x = \ketbra{\text{L}}{\text{R}} + \ketbra{\text{R}}{\text{L}}$, $\tau_z = \ketbra{\text{L}}{\text{L}} - \ketbra{\text{R}}{\text{R}}$, $\{\sigma_i\}_{x,z}$ the Pauli operators acting on the charge and spin degrees of freedom respectively and $\Omega, \omega, \phi$ the drive amplitude, frequency and phase. Here, we set $\varepsilon = 0$ as one would do in an experiment in order to be first-order insensitive to charge noise, and $\phi = 0$ as we want to perform an X gate.

\begin{figure}
    \centering
    \includegraphics[width=0.8\linewidth]{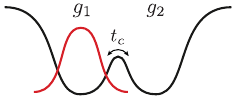}
    \caption{Tunnel-coupling based homogenisation. We consider two dots with different $g$-factor. As explained in \cref{sec:general_idea}, increasing the transfer rate $t_c$ allows an electron to acquire an effective $g$-factor equal to the average of that of the two dots. While this method does not allow for the reduction of the number of frequencies, it allows for a greater tunability as explained in \cref{app:increased_tun_tc}.}
    \label{fig:tunnel_schematic}
\end{figure}

By lowering the barrier between the dots, the electron will start oscillating (with a frequency $t_c$) between the charge states of the two dots $\ket{\text{L}}$ and $\ket{\text{R}}$ with frequencies $\omega_{q,1}$ and $\omega_{q,2}$ respectively (see \cref{fig:tunnel_schematic}). As in \cref{sec:general_idea}, we expect that sending a drive at the average frequency $\bar{\omega}_q = (\omega_{q,1}+\omega_{q,2})/2$ should result in an $X$ gate. Note that for the qubit to come back to its initial position at the end of the gate, the following condition must be satisfied,
\begin{align}
    t_c = 2p\Omega, ~~ p \in \mathbb{N}
    \label{eq:integer_tc}
\end{align}

\subsection{Numerical simulation}

\begin{figure}
    \centering
    \includegraphics[width=\linewidth]{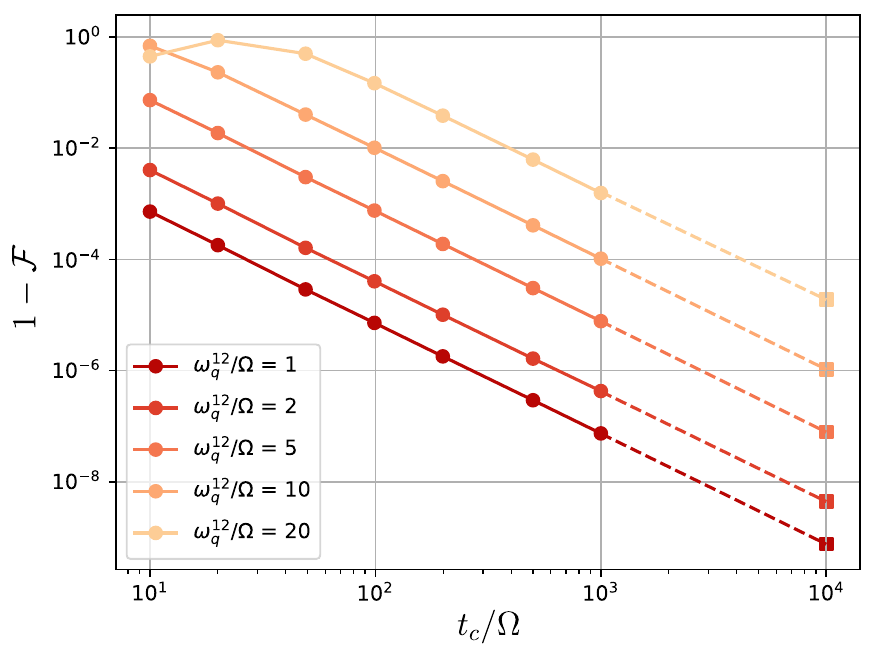}
    \caption{Infidelity of the tunnel-coupling based protocol. It is computed as the infidelity of implementing an $X$ gate when driving a qubit oscillating between two quantum dots with frequencies $\omega_{q,1}$ and $\omega_{q,2}$ respectively with a single driving tone at $\omega=\bar{\omega}_q$. The data is plotted against $t_c/\Omega$ and $\omega_q^{12}/\Omega$, where $t_c$ is the tunnel coupling and $\omega_q^{12}=\omega_{q,2}-\omega_{q,1}$. $t_c$ is always chosen such that the condition in \cref{eq:integer_tc} is satisfied. The dashed lines extend the data to $t_c/\Omega = 10^4$ which corresponds to practical values of $t_c$. The squares corresponds to the extrapolated values of the infidelity.}
    \label{fig:tunnel_coupling}
\end{figure}

In order to characterise the quality of our protocol, we plot the evolution of the fidelity (as defined in \cref{app:numerical_simulations}) between the implemented unitary $U$ and the target unitary $U_{\text{target}}$ for different values of the ratios $\omega_q^{12}/\Omega$ and $t_c/\Omega$ in \cref{fig:tunnel_coupling}. Note that the Hamiltonian used here is different from the one defined in \cref{eq:hamiltonian} as explained in \cref{app:hamiltonian_tc}. Similarly as in \cref{fig:swap_2_qubits} the infidelity naturally drops as the transfer rate increases and as $\omega_q^{12}$ decreases. Moreover, as the compute time for large $t_c$ was forbiddingly high, we obtain the last data point by linear extrapolation (dashed lines). Using values from \cref{tab:parameters}, \textit{i.e.} $t_c/\Omega = 10^4$ and $\omega_q^{12}/\Omega \gtrsim 1$, we obtain an infidelity between $10^{-7}$ and $10^{-8}$. Note that such low values will in practice be capped by other sources of noise (\textit{e.g.} charge noise). Therefore this only means that the errors induced by our protocol will not be a limiting factor.

\subsection{\label{app:increased_tun_tc} Increased tunability}

While this technique benefits from larger transfer rates and thus lower infidelities, it does not allow for a reduction of the number of distinct frequencies in the spectrum. Moreover, it requires an extra empty dot to be placed near every qubit. This allows for a larger tunability of the $g$-factors compared to that offered by the Stark shift as explained in the next section. Consider a single electron trapped in the left dot of a double quantum dot, such that its frequency is equal to $\omega_{q,1}$. In practice, the average difference in frequencies between the dots is approximately equal to $\omega_q^{12} = \Delta \omega_q = \Delta g B_0$, with $\Delta g$ the standard deviation of the $g$-factor as defined in \cref{eq:g_ou_process}. By lowering the barrier, we showed previously that the electron obtains an average frequency equal to $\bar{\omega}_q = (\omega_{q,1}+\omega_{q,2})/2$. This means that we managed to tune the frequency by an amount $\bar{\omega}_q - \omega_{q,1} = \Delta \omega_q/2$. Usually, the frequency of an electron can be tuned using the Stark shift by an amount $\delta \omega_q = 0.1\Delta \omega_q < \Delta \omega_q/2$. Such increased tunability could be leveraged when needed, in order to increase the distance between qubit frequencies and thus reduce crosstalk.

\section{Crosstalk suppression via synchronisation} \label{app:crosstalk_sync}

In the application of our protocol to the 2$\times$N we present in \cref{subsec:two_by_n}, we send global pulses aiming to drive a subset of the qubits, and wish to minimise their impact on all other qubits. This is in general enforced by maximising the frequency spacing between drives and non-targetted qubits (thereby minimising off-resonant effects). Here we show that additionally tuning the drive strength $\Omega$ can help further mitigate these crosstalk effects. Namely, a clever choice of $\Omega$ can allow for the synchronisation of the oscillations of the electrons, such that the non-target qubits perform a $2\pi$ rotation (identity gate) while the target qubits perform a $\pi$ rotation ($X$ gate). 

Let us first focus on the case where only one target frequency and one non-target frequency are present in the frequency spectrum. This essentially describes our shuttling-based applications from \cref{subsec:two_by_n} (although in general there subsists a small dispersion $\sigma$ around these main frequencies). In this case, one can equate the times taken by one $X$ gate on the target qubits and a $2p\pi$  ($p\in\mathbb{N}$) rotation of the non-target qubits by enforcing the condition,
\begin{equation}
    \frac{\pi}{\Omega} = \frac{2p\pi}{\sqrt{\Omega^2+(\omega_q^{12})^2}},
\end{equation}
with $\omega_q^{12}$ the difference between the effective target and non-target frequencies. 
The target qubits theoretically oscillate at a frequency $\Omega$, while off-resonant effects induce oscillations of the non-target qubits at frequency $\sqrt{\Omega^2+(\omega_q^{12})^2}$ as inferred from the Rabi model. This results in setting:
\begin{equation}
    \Omega = \frac{\omega_q^{12}}{\sqrt{4p^2-1}}, ~~ p\in\mathbb{N}
    \label{eq:synchro_condition}
\end{equation}
By reducing the spectrum to only one target and one non-target frequencies, setting the above condition \textit{totally} removes any crosstalk infidelity as non-target qubits exactly perform an identity gate while target qubits perform an $X$ rotation.

When using the binning scheme, it is in general not possible to reduce the frequency spectrum to only two frequencies. In this case, we can generalise the above condition as follows (see Section II of \cite{Fayyaz_2023} with a bin width $2\delta\omega_q$):
\begin{equation}
    \Omega = \frac{\delta \omega_q}{p}, ~~ p\in\mathbb{N}
    \label{eq:bins_condition}
\end{equation}
where $p$ is a free parameter that can be adjusted according to the achievable values of $\Omega$ and $\delta \omega_q$. This is a generalisation of \cref{eq:synchro_condition}, where the existence of only two frequencies in the frequency spectrum permitted the complete removal of crosstalk effects. Here the extension to more than two frequencies only allows for a partial reduction of the crosstalk.

Note that in both cases $\Omega \leq \omega_q^{12}$, meaning that if the frequencies are close to each other, slower gates will be obtained.

\section{Estimation of the right parameter regime} \label{app:param_estimation}

\subsection{Shuttling-based scheme}

In \cref{subsec:problem_statement}, we stated the values of experimental parameters which \textit{a posteriori} led to a high performance of our shuttling-based protocol. In this section, we further justify this choice with some analytical arguments.

Firstly, our shuttling-based homogenisation performs best at low frequency dispersion, justifying to work in the regime $\Delta g=10^{-3}g_0$ and $B_0=0.1$ T. Second, the dispersion of target frequencies around the driving frequency (due to finite $\sigma$) means that target qubits are not exactly driven at resonance. This undesired effect is minimised at large $\Omega$, which is also synonym of faster gates. This is also observed in the second panel of \cref{fig:shuttling_proc_inf}. For this reason, we set $\Omega=5$ MHz.

Let us now evaluate the target values of the last key experimental parameters enabling the full power of our protocol \textit{i.e.} the required $g$-factor shift $G$ between targets and non-targets (for the 2$\times$N case), the shuttling length $d$ and the shuttling speed $v$ (for both architectures). When qubits are shuttled back and forth in a line, $d$ is the one way distance; when qubits are shuttled in a loop, $d$ is the loop length.  

We will first focus on the 2$\times$N case where a uniformly frequency shift $GB_0$ is applied between the targets and non-targets. In this scenario $\sigma >0$. We will estimate the above parameters by distinguishing two main contributions to the infidelity: the slightly off-resonant driving of the homogenised target qubits and the crosstalk with the non-target qubits. We wish to bring each of these contributions below $2\times10^{-3}$ for at least $95\%$ of the qubits, so as to be comfortably below the surface code threshold. If the locations of the $5\%$ outliers are known, the error they generate can be mitigated by suitable QEC protocols as erasure errors \cite{Sahay_2023}.

The first contribution reads:
\begin{equation}
    I_1 = \frac{1}{1 + \left(\frac{\Omega}{2\sigma B_0}\right)^2}
\end{equation}
from the Rabi model, where we use $2\sigma B_0$ for the frequency mismatch between the drive and a given target qubit so as to cover $95\%$ of the qubits. $\sigma$ is the standard deviation of the homogenised $g$-factors
\begin{equation}
    \bar{g} = \frac{1}{d}\int_{0}^{d}g(x)\mathrm{d}x.
\end{equation}
In a first approximation, we can write
\begin{equation}
    \sigma = \frac{\Delta g}{\sqrt{d/\lambda}}
\end{equation}
where $\lambda$ is the coherence length of the $g$-factor landscape. This equation simply states that the standard deviation of the average of independent random variables decreases as $1/\sqrt{N}$, where $N$ is the number of samples. This leads to:
\begin{equation}
    I_1 = \frac{1}{1 + \frac{\Omega^2}{4\Delta\omega_q^2} \frac{d}{\lambda}}
\end{equation}
with $\Delta\omega_q=\Delta g B_0$. By setting $I_1=2\times 10^{-3}$ we find a minimum shuttling distance $d\geq 14 \;\mu$m. From this we can deduce the minimum shuttling speed enabling the exploration of such a distance during the gate time $T=\pi/\Omega$:
\begin{equation}
    v = \frac{\Omega d}{\pi}
\end{equation}
We deduce $v \geq 23\text{ m/s}$. We can now evaluate the infidelity arising from the crosstalk with non-target qubits:
\begin{equation}
    I_2 = \frac{1}{1+\left(\frac{GB_0}{\Omega}\right)^2}
\end{equation}
We here neglect the dispersion $\sigma$ of the non-target qubits around their mean $g$-factor as it is negligible compared to $G$. We here find $G\geq 160$ MHz. Note that for simplicity, we do not describe here the crosstalk reductions offered by the technique of \cref{app:crosstalk_sync}. This would permit for lower values of $G$.

For the looped-pipeline architecture, when no frequency shift is applied, we are mainly limited by the off-resonant driving of the targets (as there a no non-target qubits). The derivation of $I_1$ still holds in this case, which allow us to deduce a similar lower bound on the shuttling distance $d$ and shuttling speed $v$. In the case where the magnetic field gradient can be tuned so as to annihilate the frequency dispersion $\sigma$, the infidelity is only limited by the performance of our homogenisation protocol (which we neglected in the previous case). Indeed, $I_1$ vanishes as $\sigma=0$ (all targets are resonantly driven). As for $v$ and $d$, their values can be inferred from the top panel of \cref{fig:shuttling_proc_inf}, which describes a driving at exact resonance with the homogenised target frequency. At $\Omega=5$ MHz, the resulting infidelity (from imperfect homogenisation) never exceeds $0.4\%$, meaning no value of $d$ and no value of $v\geq 10$ m/s would be an impediment to surface-code-enabled error correction.

Note however that these models are highly simplified for the ease of parameter estimation, using worst-case approximations. Consequently, although the first parameter regime we considered in \cref{sec:applications} (\cref{fig:infid_2xN}) does not rigorously respect the conditions we here set, it still yields desirable performances.

\subsection{Binning scheme}

We proceed similarly for the binning scheme. Its performance is maximised at large interbin spacing, which is given by $2\delta\omega_q=2\Delta gB_0/10$. It is thus optimal to work at higher $g$-factor spread and higher magnetic field in this case: $\Delta g=10^{-2}g_0$ and $B_0=1$ T. Another reason for this choice is that enforcing \cref{eq:bins_condition} for crosstalk reduction (which is crucial here) implies $\Omega \leq \delta\omega_q$. $\delta\omega_q$ must therefore be kept relatively high to ensure that gates are fast. With the above choice of parameters, we obtain $\delta\omega_q=30$ MHz, which would not be a bottleneck. Besides, the previous analysis leading to $G \geq 160$ MHz is still valid here.

\subsection{Hybrid shuttling-binning scheme}
\begin{figure*}[!]
    \includegraphics[width=\linewidth]{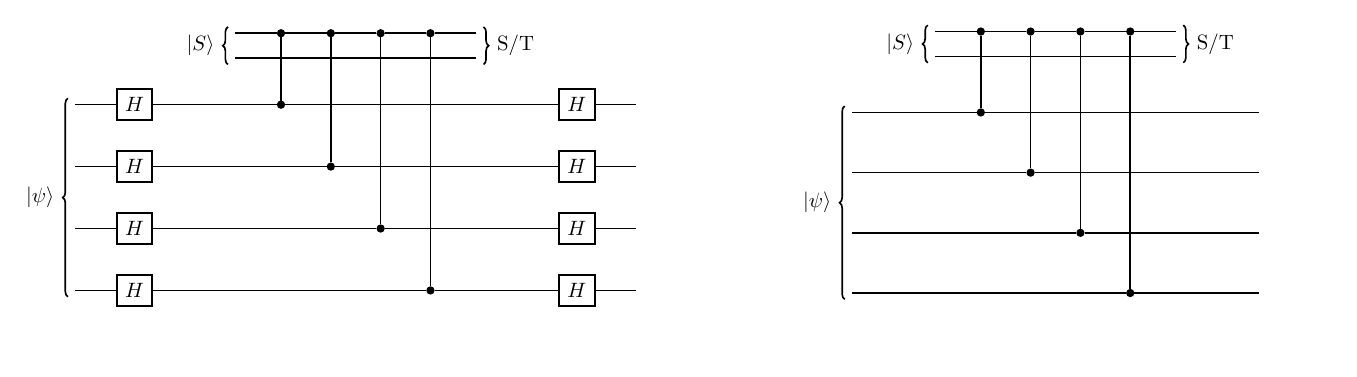}
    \caption{Stabiliser circuits for silicon spin qubits. The ancilla qubit is encoded in the singlet-triplet basis while the data qubits are of the Loss-DiVincenzo type. (left) X-type stabilizer circuit in which the ancilla qubits can be ignored while manipulating the data qubits.(right) Z-type stabiliser circuit in which we are not addressing the data qubits. }
    \label{fig:stab_cycle}
\end{figure*}

We briefly mention here that one further protocol we envisioned was to combine the shuttling and binning schemes. Once qubits are shuttled, rather than targetting them at a mean slightly off-resonant frequency $\omega=GB_0$, one could envisage to place the post-shuttling effective frequencies into bins and resonantly target them with multiple driving tones. While this two-step frequency factorisation protocol sounds powerful on paper, it would not lead to desirable fidelities as the shuttling and binning schemes are not operated in the same parameter regime (they respectively work best at low and high $\Delta\omega_q$).

\section{\label{app:stab_cycle} Stabiliser cycles for silicon spin qubits}

The goal of this section is to show that for stabiliser cycles, we can ignore the ancilla qubits when globally targetting the data qubits. 
For silicon spin qubits, the X and Z stabiliser circuits can be implemented as in \cref{fig:stab_cycle}, making use of the native CZ gate. Note that we use a singlet state for the ancilla (instead of a single qubit prepared in $\ket{+}$). Indeed, singlet initialisation and singlet/triplet measurement are permitted by silicon spin qubits.
Crucially, this removes the need for single-qubit gates on the ancilla. Therefore, state preparation (measurement) of the ancilla can be performed after (before) the first (last) global Hadamard gates as illustrated in \cref{fig:stab_cycle}. As such, the ancilla qubits do not need to be considered as non-target qubits when we address the data qubits.

%

\end{document}